\documentclass[twocolumn,showpacs,aps,amsmath,amssymb,superscriptaddress,prb]{revtex4-1}
\usepackage{amsfonts}
\usepackage{amsmath}
\usepackage{bm}
\usepackage{setspace}
\usepackage{graphicx}
\usepackage{overpic}
\usepackage{color}
\usepackage{multirow}
\usepackage{url}
\usepackage{array}
\usepackage{makecell}
\usepackage{pifont}
\usepackage{amssymb}
\usepackage{simplewick}
\newcommand{\PreserveBackslash}[1]{\let\temp=\\#1\let\\=\temp}
\newcolumntype{C}[1]{>{\PreserveBackslash\centering}p{#1}}
\newcolumntype{R}[1]{>{\PreserveBackslash\raggedleft}p{#1}}
\newcolumntype{L}[1]{>{\PreserveBackslash\raggedright}p{#1}}
\newcommand{\Sec}[1]{Sec.\,\ref{#1}}

\newcommand{\nl}{\nonumber \\}
\newcommand{\be}{\begin{equation}}
\newcommand{\ee}{\end{equation}}
\newcommand{\bea}{\begin{eqnarray}}
\newcommand{\eea}{\end{eqnarray}}

\newcommand{\Eq}[1]{Eq.\,(\ref{#1})}
\newcommand{\Eqs}[1]{Eqs.\,(\ref{#1})}
\newcommand{\Sch}{Schr\"{o}dinger\ }
\newcommand{\la}{\langle}
\newcommand{\ra}{\rangle}

\newcommand{\barxi}{\bar{\xi}}

\newcommand{\iChEM}{{\it i}{\rm ChEM}}%
\newcommand{\iLsb}{\mathcal{L}_{_{\rm SB}}^{\rm I}}
\newcommand{\tLsb}{\tilde{\mathcal{L}}^{\,\rm I}}
\newcommand{\itrhos}{\tilde{\rho}_{_{\rm S}}^{\, \rm I}}

\newcommand{\bpsi}{\bar{\psi}}

\newcommand{\baeta}{\bar{\eta}}
\newcommand{\btheta}{\bar{\theta}}

\newcommand{\trb}{{\rm tr}_{_{\rm B}}}
\newcommand{\trs}{{\rm tr}_{_{\rm S}}}
\newcommand{\brho}{\bar{\rho}}
\newcommand{\brhos}{\bar{\rho}_{_{\rm S}}}

\newcommand{\rhos}{\rho_{_{\rm S}}}
\newcommand{\rhob}{\rho_{_{\rm B}}}
\newcommand{\barrhos}{\bar{\rho}_{_{\rm S}}}
\newcommand{\trho}{\tilde{\rho}}
\newcommand{\trhos}{\tilde{\rho}_{_{\rm S}}}

\newcommand{\hlambda}{\lambda^{\frac{1}{2}}}
\newcommand{\mhlambda}{\lambda^{-\frac{1}{2}}}
\newcommand{\huz}{\hat{U}_{_{\rm B}}^{(0)}}

\newcommand{\htb}{\hat{\tilde{U}}_{_{\rm B}}}

\newcommand{\Hs}{H_{_{\rm S}}}
\newcommand{\Hb}{H_{_{\rm B}}}
\newcommand{\hc}{\hat{c}}
\newcommand{\hF}{\hat{F}}
\newcommand{\hd}{\hat{d}}

\newcommand{\gm}{g^{-}}
\newcommand{\gp}{g^{+}}
\newcommand{\Gm}{\mathcal{G}_{\bm m}^{\bm -}}
\newcommand{\Gp}{\mathcal{G}_{\bm n}^{\bm +}}

\newcommand{\Cm}{C^{-}}
\newcommand{\Cp}{C^{+}}

\newcommand{\mD}{\mathcal{D}}

\newcommand{\Xu}{X^+}
\newcommand{\Xd}{X^-}

\newcommand{\tgm}{\tilde{g}^-}
\newcommand{\tgp}{\tilde{g}^+}

\begin{document}

\title{Stochastic Equation of Motion Approach to Fermionic Dissipative Dynamics. I. Formalism}

\author{Lu Han}
\affiliation{Hefei National Laboratory for Physical Sciences at the Microscale \&
	Synergetic Innovation Center of Quantum Information and Quantum Physics,
	University of Science and Technology of China, Hefei, Anhui 230026, China}

\author{Arif Ullah}
\affiliation{Hefei National Laboratory for Physical Sciences at the Microscale \&
	Synergetic Innovation Center of Quantum Information and Quantum Physics,
	University of Science and Technology of China, Hefei, Anhui 230026, China}

\author{Yun-An Yan}
\affiliation{School of Physics and Optoelectronic Engineering, Ludong University, Shandong 264025, China}

\author{Xiao Zheng} \email{xz58@ustc.edu.cn}
\affiliation{Hefei National Laboratory for Physical Sciences at the Microscale \&
	Synergetic Innovation Center of Quantum Information and Quantum Physics,
	University of Science and Technology of China, Hefei, Anhui 230026, China}

\author{YiJing Yan}
\affiliation{Hefei National Laboratory for Physical Sciences at the
	Microscale \& \iChEM, University of Science and Technology of China, Hefei, Anhui
	230026, China}

\author{Vladimir Chernyak}
\affiliation{Hefei National Laboratory for Physical Sciences at the Microscale \&
	Synergetic Innovation Center of Quantum Information and Quantum Physics,
	University of Science and Technology of China, Hefei, Anhui 230026, China}
\affiliation{Department of Chemistry, Wayne State University, 5101 Cass Avenue, Detroit, MI 48202}

%\date{\today}
\date{Submitted on December~10, 2019}

\begin{abstract}

  In this work, we establish formally exact stochastic equations of motion (SEOM)
  theory to describe the dissipative dynamics of fermionic open systems.
  The construction of the SEOM is based on a stochastic decoupling of
  the dissipative interaction between the system and fermionic environment,
  and the influence of environmental fluctuations on the reduced system dynamics
  is characterized by stochastic Grassmann fields.
  Meanwhile, numerical realization of the time-dependent Grassmann fields
  has remained a long-standing challenge.
  To solve this problem, we propose a minimal auxiliary space (MAS) mapping scheme,
  with which the stochastic Grassmann fields are represented
  by conventional c-number fields along with a set of pseudo-levels.
  This eventually leads to a numerically feasible MAS-SEOM method.
  The important properties of the MAS-SEOM are analyzed
  by making connection to the well-established time-dependent perturbation theory
  and the hierarchical equations of motion (HEOM) theory.
  The MAS-SEOM method provides a potentially promising approach for
  accurate and efficient simulation of fermionic open systems at ultra-low temperatures.

\end{abstract}

%
%\pacs{73.22.-f, 71.27.+a, 72.15.Qm, 73.63.-b}

%%%pacs:
%%%73.22.-f	  Electronic structure of nanoscale materials and related systems
%%%71.27.+a   Strongly correlated electron systems
%%%72.15.Qm	  Scattering mechanisms and Kondo effect
%%%73.63.-b	  Electronic transport in nanoscale materials and structures

\maketitle

\section{Introduction} \label{sec:intro}

The dissipative dynamics of a quantum system embedded in a macroscopic environment
is a fundamental problem in modern physics and chemistry.
To address this problem, a number of formally exact quantum dissipation
theories (QDTs) have been established over the past two decades.\cite{Wei12,Bre16021002,Veg17015001}

One of the most popular QDTs is the hierarchical equations of motion (HEOM) theory proposed first by Tanimura and Kubo \cite{Tan89101,Tan906676}
and extended later by many authors.\cite{Yan04216,Xu05041103,Jin08234703,Li12266403,Tan14044114,Liu14134106,Sch18235429}
%%%Xiao: too many references for HEOM?
%
Accurate and efficient numerical schemes have been developed for the HEOM,\cite{Shi09084105,Zhe09164708,Hu10101106,Hu11244106,Str122808,
Tsu153859,Hou15104112,Ye17074111,Erp18064106,Han18234108,Rah19244104,Cui19024110,%%% The following are references on GPU.
Kre112166,Kre144045,Tsu153859,Kra181779} %%%Xiao: cite the GPU-implentation papers
which have led to a wide range of applications
including quantum-state evolution,\cite{Ish053131,Zhe08093016,Zhe09124508,Che153110,Tan15224112}
optical spectroscopy,\cite{Tan9466,Ish06084501,Che11194508,Zhu115678,Hei12023018}
energy and heat transfer,\cite{Ish0917255,Ish09234111,Kre112166,Kre144045,Ye14165116,Kat15064107,Zha16204109,Son17064308}
charge transfer and transport,\cite{Zhe08184112,Shi09164518,Tan10214502,Zhe13086601,Wan13035129,Har13235426,Har15085430,Sch16201407}
exciton dynamics in realistic molecular aggregates,\cite{Sch151,Ke151741,Zan175105}
manipulation of qubits,\cite{Dij10250401,Ma12062323,Hou15104112,Hon18510}
quantum phase transition,\cite{Hou14045141}
and strongly electron correlation effects.\cite{Wan14084713,Wan16125114,Wan16034101,Wan16154301,Ye16608}
%%%Xiao: too many papers cited, consider to remove some not-so-high-impact papers
%
Despite its great success, the applicability of HEOM is often limited by the considerable cost of
computer memory, which tends to increase drastically with the lowering of temperature.\cite{Hou15104112,Han18234108,Cui19024110,Cui192}
This is because the construction of HEOM is based on unraveling the non-Markovian environmental memory.
A large memory basis set is usually needed to achieve an accurate unraveling at low temperatures,
which inevitably requires a large amount of computer memory to store all the auxiliary density operators.

%For instance, non-Markovian effects emerge in the long-time dynamics of
%strongly correlated quantum impurity systems at low temperatures,
%which normally result in the slow convergence of HEOM based on
%conventional schemes for decomposing the reservoir correlation functions.\cite{Cui19024110}
%

Apart from the deterministic HEOM theory, there is
another class of QDTs which adopts a stochastic framework.
Since a stochastic process can be realized via
mutually independent trajectories, the numerical
implementation of a stochastic QDT is memory-friendly and highly parallelizable.

The stochastic QDTs have been well established for bosonic dissipative systems.
For instance, the quantum state diffusion (QSD) theory using a wavefunction description
pioneered by Di\'{o}si,\cite{Dio882885} and Gisin and Percival\cite{Gis92315,Gis925677,Per99}
has been extended to capture the non-Markovian effects
by Strunz \emph{et~al.}\cite{Str9625,Dio97569,Dio981699,Str991801,Yu9991,Sue14150403,Lin17180401}
and many others.\cite{Kle95224,Yu04062107,Jin10240403,Tam18030402}
Alternatively, using a density operator description,
Kubo has developed a stochastic Liouville equation for quantum systems
as early as in 1963.\cite{Kub63174}
%Cao, Ungar and Voth have proposed to use Gaussian noises to sample the
%fluctuating environmental force for quantum dissipative systems.\cite{Cao964189}
Later, Stockerger and Grabert\cite{Sto02170407} and Shao \cite{Sha045053} have independently
established the formally exact stochastic equations of motion (SEOM) theory.
In contrast to the HEOM theory, the SEOM formalism does not involve the unraveling of environmental memory.
Instead, stochastic fields are introduced to represent the environmental fluctuations\cite{Cao964189}
and to decouple the dynamics of system and the environment.

For bosonic environments, the stochastic fields are conveniently realized by c-number random noises.
Yan and Shao \emph{et~al.} have proved the formal equivalence between the HEOM and the SEOM theories
for bosonic environments,\cite{Yan04216,Yan161,Yan19164110}
as well as the equivalence between the HEOM and the non-Markovian QSD theories.\cite{Yan18042126}
They have also applied the SEOM to study the dynamics of a spin-boson model at zero temperature.\cite{Yan15022121,Yan161}
Hsieh and Cao have proposed a unified SEOM formalism for various types of environments.\cite{Hsi18014103,Hsi18014104}
Besides, hybrid stochastic and hierarchical equations of motion (sHEOM) formalisms
have also been established.\cite{Zho05334,Zho08034106,Moi13134106,Zhu13095020}

In contrast to the bosonic counterparts, the stochastic QDTs for
fermionic open systems have remained far from mature.
Historically, the pioneering works can be traced back to the
early attempts of Barnett \emph{et~al.},\cite{Bar82172,Bar831469,Bar8313}
Applebaum \emph{et~al.},\cite{App84473,App8717} and Rogers,\cite{Rog87353}
who have derived stochastic dynamical equations to describe fermionic Brownian motion.
In recent years, the non-Markovian QSD theory has been extended to address fermionic environments
by Yu \emph{et~al.}\cite{Zha12032116,Che14052104,Zha17121} and others.\cite{Che13052108,Sue151408}
Moreover, the SEOM for fermionic environments has been formally established.\cite{Hsi18014103}
Since the creation and annihilation operators of fermions satisfy the
anti-commutation relationship, the random fluctuations in a fermionic environment
are to be represented by stochastic Grassmann-number (g-number) fields.\cite{Alt2010}

It should be pointed out that, all the previous efforts on the stochastic formulations
of fermionic open systems ended up with derivations of deterministic dynamical equations.\cite{Che13052108,Sue151408,Hsi18014103}
The inability to implement such stochastic formulations is due to intrinsic
difficulties originating from the nature of Grassmann variables.

Any conventional variable, continuous or discrete, can be viewed as a
function in some configuration space.
Alternatively, as it is done in algebraic geometry, one can view functions as the basic object,
with the underlying space being retrieved once the algebra of functions is given.
The fermionic SEOM theory can be formulated in terms of functions which satisfy the Grassmann algebra,
while completely avoiding the notion of points.
However, the underlying space of points associated with the Grassmann variables does not exist.
This makes the applications of Monte Carlo sampling of high-dimensional or path integrals
to the integrals over Grassmann variables highly problematic.\cite{Ber1966}
Consequently, the implementation of stochastic g-numbers is rather difficult in practice.
For instance, a set of $N$ matrices with the size of $2^{N} \times 2^{N}$
are needed to represent $N$ g-numbers.
Here, $N$ could be the number of time steps considered in a dynamical process.
Apparently, as $N$ increases, the size of matrix will soon exceed the limit of current computers.

Recently, we have proposed a mapping strategy for the stochastic g-number fields,
based on which a numerically feasible SEOM approach has been developed.\cite{Han19050601}
This has enabled the stochastic simulation of fermionic dissipative dynamics.
Although the main idea has been given in Ref.~\onlinecite{Han19050601},
a lot of details about the practical SEOM approach are yet to be elucidated.
To this end, we give a comprehensive account of the fermionic SEOM theory
and related numerical approach in a series of two full papers.
In this paper (paper I), we shall focus on the analytic formalism of the SEOM theory.
We will not only give detailed derivation of the SEOM theory,
but also provide important insights into the mapping strategy that is
essential for achieving a numerically feasible SEOM approach.
In a succeeding paper (paper II),\cite{Ari19}
we will elaborate on the numerical aspects
of the SEOM approach with numerical demonstrations.

The remainder of this paper is organized as follows.
In \Sec{sec:SDE_fermion} we utilize a stochastic decoupling scheme to
derive a formally exact SEOM for open fermionic systems.
In addition, the equivalence between the fermionic SEOM and HEOM formalisms
is established by using the Ito calculus.
In \Sec{sec:auxiliary_space} we elaborate on the mapping scheme
for the stochastic g-number fields,
with which the formally exact yet numerically unfeasible SEOM
is converted into a numerically feasible SEOM.
In \Sec{sec:limitation} we assess the exactness or non-exactness of the SEOM
with two different analytic approaches.
Finally, we give concluding remarks in \Sec{sec:summary}.

\section{A stochastic Framework for open fermionic systems}\label{sec:SDE_fermion}

\subsection{Fermionic dissipative systems}

In this work, we consider a generic system coupled to fermionic environments
(such as electron reservoirs).
The total system (system plus environment) is described by the Hamiltonian
\be \label{H_t}
H_{_{\rm T}} = H_{_{\rm S}} + H_{_{\rm B}} + H_{_{\rm{SB}}}.
\ee
Here, $H_{_{\rm S}}$, $H_{_{\rm B}}$ and $H_{_{\rm{SB}}}$ are the Hamiltonian
of the system, the Hamiltonian of the bath environment,
and the system-bath interaction, respectively.
In particular, for a single-level system in contact with a non-interacting fermion bath,
we have
%\be \label{Hb-Hsb}
$H_{_{\rm B}} = \sum_{j}\epsilon_{j}\, \hd_{j}^{\,\dag}\hd_{j}$
and  $H_{_{\rm{SB}}} = \hc^\dag \hat{F} + \hat{F}^\dag\hc$,
where $\hc\,(\hc^\dag)$ and
$\hd_{j}\,(\hd_{j}^{\,\dag})$ are the annihilation (creation) operators
acting on the system level and $j$th single-particle state of the fermion bath.
$\hat{F} = \sum_{j} t_{j} \hd_{j}$, where $t_{j}$ is the system-bath coupling strength.

The dynamics of the total system is determined by the Schr\"odinger equation,
\be \label{def:total-eom}
i {\dot{\rho}_{_{\rm T}}} = [H_{_{\rm T}},\, \rho_{_{\rm T}}],
\ee
where $\rho_{_{\rm T}}$ is the density matrix of the total system.
In this paper, we use the atomic units ($e = \hbar \equiv 1$) and $k_{\rm B}\equiv 1$.
The straightforward computation of \Eq{def:total-eom} is intractable,
because the fermion bath includes an infinite degrees of freedom (DOF) in the $\Hb$.

For a quantum dissipative system, we are interested in its reduced dynamics as well as its static and dynamical properties.
The dissipative dynamics is characterized by the system reduced density matrix
$\brho_{_{\rm S}}(t) = {\rm tr}_{_{\rm B}}[\rho_{_{\rm T}} (t)]$.
In the following, we establish a rigorous stochastic formalism for the evaluation of $\brho_{_{\rm S}}(t)$.
For simplicity, the environment consists of a single fermion bath.
As will be shown later, the derivation is easily extended to systems coupled to multiple fermion baths.

\subsection{Fermionic coherent states and Grassmann algebra}

For a single-level system, $\hc^\dag$ and $\hc$ are respectively associated with
the fermionic creation and annihilation operator of system level.
The fermionic coherent states, $|\psi\ra$ and $\la\bpsi|$, are defined as
\begin{align}
|\psi \ra & \equiv e^{-\psi \hc^\dag} |0 \ra
= \left(1- \psi \hc^\dag \right) |0\ra = |0\ra - \psi |1\ra\label{def-cs-psi} \\
\la \bpsi | & \equiv \la 0 | \, e^{\bpsi \hc}
= \la 0 | \left(1+\bpsi \hc \right) = \la 0 | - \la 1 | \bpsi,   \label{def-cs-bpsi}
\end{align}
where $|0\ra$ and $|1\ra$ are the Fock states with the particle occupation
number being zero and one, respectively.
$\bpsi$ and $\psi$ are g-numbers. They satisfy the anti-commutation relation of
$\bpsi\psi =-\psi\bpsi$, and hence $(\bpsi)^2  = (\psi)^2 = 0$.
It can be easily verified that $|\psi \ra$ and $\la \bpsi |$ are
the eigenstates of operators $\hc$ and $\hc^\dag$,
with the eigenvalues $\psi$ and $\bpsi$, respectively.
%
%\begin{align}
\be
\hc |\psi \ra = \psi |\psi \ra \quad  {\rm and} \quad  \la \bpsi | \hc^\dag = \la \bpsi | \bpsi.
\ee
%\end{align}
In addition, we also have
\be
\hc^\dag |\psi \ra = -\frac{\partial}{\partial \psi} |\psi \ra
\quad {\rm and} \quad
\la \bpsi | \hc = \frac{\partial}{\partial \bpsi} \la \bpsi |.
\ee
The inner product of two fermionic coherent states is
\be
\la \bpsi | \psi \ra = 1 + \bpsi\psi = e^{\bpsi \psi}.
\ee
All the fermionic coherent states form an overcomplete set,
and thus the identity operator in the system's Hilbert space,
$\hat{I}$, is expressed as
\be
  \hat{I} = |0\ra \la 0| + |1\ra \la 1| = \int d\bpsi \, d\psi e^{-\bpsi \psi}|\psi \ra \la \bpsi|.
\ee

The above definitions and relations can be extended to
a system with $N_\nu$ DOF. In such a case, define
\begin{align}
|\psi \ra & \equiv |\{ \psi_\nu\} \ra
= e^{-\sum_{\nu=1}^{N_\nu} \psi_\nu \hc_\nu^\dag} |0\ra
= |0\ra - \sum_{\nu=1}^{N_\nu} \psi_\nu |1_\nu \ra,  \label{def-cs-psi-nu} \\
\la \bpsi | & \equiv \la \{ \bpsi_\nu\} |
= \la 0 | e^{\sum_{\nu=1}^{N_\nu} \bpsi_\nu \hc_\nu}
= \la 0 | - \sum_{\nu=1}^{N_\nu} \la 1_\nu | \bpsi_\nu. \label{def-cs-bpsi-nu}
\end{align}
Here, $|0\ra \equiv |0_1,\cdots,0_{N_\nu}\ra$ is the anti-symmetrized
vacuum state of the system, and $|0_\nu\ra$ and $|1_\nu\ra$
are the Fock states associated with the $\nu$th level. The inner product is
evaluated by
\be
\la \bpsi | \psi \ra = 1 + \sum_{\nu=1}^{N_\nu} \bpsi_\nu \psi_\nu
= e^{\sum_{\nu=1}^{N_\nu} \bpsi_\nu \psi_\nu}.
\ee
The identity operator is reproduced by
\be
\hat{I} = \int \prod_{\nu=1}d\bpsi_\nu d\psi_\nu
e^{-\sum_{\nu=1}\bpsi_\nu \psi_\nu}|{ \psi}_\nu \ra \la { \bpsi}_\nu|.
\ee

\subsection{Stochastic decoupling of system and environment dynamics}

For a fermionic open system, the stochastic decoupling of system and bath
is done by following a procedure similar to that adopted for a bosonic open system.\cite{Sha045053,Sch151}
The evolution of the density matrix of the total system is formally described by
\be\label{def:tot-evolution}
\rho_{_{\rm T}} (t) = \hat{U}_{_{\rm T}} (t,\,0) \rho_{_{\rm T}}(0) \,\hat{U}_{_{\rm T}}^\dag(t,\,0).
\ee
where $\hat{U}_{_{\rm T}}(t,\,0)$ and $\hat{U}_{_{\rm T}}^\dag(t,\,0)$
are the forward and backward quantum evolution operators of the total system, respectively.
Denote $\mathcal{U}_{_{\rm T}}(t,\, 0)$ and $\mathcal{U}_{_{\rm T}}^\dag(t,\, 0)$ as the
corresponding quantum propagators in the fermionic coherent state representation,
and discretize the time domain by an infinitesimal time interval
$\Delta t = t_{i+1} - t_{i} = t/N$ with $t_0=0$ and $t_N=t$.
The quantum propagator is expressed as follows.
\begin{align} \label{pro:total}
\mathcal{U}_{_{\rm T}} (t,\,0) &= \la\, {\bar{\psi}_{_N} \bar{\theta}_{_N}}\,|\hat{U}_{_{\rm T}} (t,\,0)|\, \psi_{_0}\theta_{_0}\,\ra \nl
& =\lim \limits_{N \rightarrow \infty} \int \mathcal{D} \bm\psi  \mathcal{D} \bm\theta \prod\limits_{i=0}^{N-1}
 \la {\bar{\psi}_{i+1}\bar{\theta}_{i+1}}\,|\hat{U}_{_{\rm T}} (t_{i+1},\,t_i)|\,\psi_i \theta_i \ra \nl
&= \lim \limits_{N \rightarrow \infty}\int \mathcal{D} \bm\psi  \mathcal{D} \bm\theta
 \prod\limits_{i=0}^{N-1} \mathcal{U}_{_{\rm T}} (t_{i+1},\,t_i).
\end{align}
Here, the g-numbers $\{\,\bpsi_{i},\,\psi_i, \btheta_{i},\,\theta_{i}\,\}$
are eigenvalues of the system and bath operators
$\{\hlambda\hat{c}^\dag,\, \hlambda\hat{c},\, \mhlambda\hat{F}^\dag,\, \mhlambda\hat{F}\}$,
and $\lambda$ is a reference energy that could take any positive value $(\lambda > 0)$.
$\mathcal{D} \bm\psi  \mathcal{D} \bm\theta \equiv \prod\limits_{i=1}^{N-1}
\frac{1}{(\Delta t)^2}\, d\bar{\psi}_i d\psi_i \, d\bar{\theta}_i d\theta_i \, e^{-(\bar{\psi}_i\psi_i+\bar{\theta}_i\theta_i)\Delta t}$
is the metric of the integral.
The overcompleteness of fermionic coherent states gives rise to the equality
\be \label{def:completeness}
\hat{I}=\frac{1}{(\Delta t)^2}\int d\bar{\psi}_i d\psi_i \, d\bar{\theta}_i d\theta_i\,
e^{-\bar{\psi}_i\psi_i \Delta t}e^{-\bar{\theta}_i\theta_i \Delta t}\,
|{\psi_i \theta_i}\,\ra \la \, {\bar{\psi}_i \bar{\theta}_i} |.
\ee

In the limit of $N \rightarrow \infty$ ($\Delta t \rightarrow 0$),
the propagator $\mathcal{U}_{_{\rm T}} (t_{i+1},\,t_i)$ can be further split into three parts,
\be \label{pro:short}
 \mathcal{U}_{_{\rm T}} (t_{i+1},\,t_{i}) = \mathcal{U}_{_{\rm S}} (t_{i+1},\,t_{i})\,
 \mathcal{U}_{_{\rm B}} (t_{i+1},\,t_{i}) \, \mathcal{U}_{_{\rm SB}} (t_{i+1},\,t_{i}).
\ee
The last part associated with the system-bath coupling, $\mathcal{U}_{_{\rm SB}}$,
is recast into the fermionic coherent state representation as
\begin{align} \label{pro:int}
\mathcal{U}_{_{\rm SB}} (t_{i+1},\,t_i)&=  \la \,{\bar{\psi}_{i+1}\bar{\theta}_{i+1}}\,|\hat{U}_{_{\rm SB}} (t_{i+1},\,t_i)|\,\psi_i \theta_i \, \ra \nonumber \\
&= \exp\{-i (\bpsi_{i+1}\,\theta_{i} + \,\btheta_{i+1}\psi_i)\, \Delta t \} \nonumber \\
&= \exp\{-i \bpsi_{i+1}\theta_{i}\,\Delta t\}\, \exp\{ -i \btheta_{i+1}\, \psi_i \,\Delta t \}.
\end{align}
The products of g-numbers corresponding to the system and bath operators can be decoupled
by using the property of Grassmann Gaussian integrals.
We have the following equalities:
\begin{align}
e^{-i \bpsi_{i+1} \theta_i \Delta t}  &= \frac{1}{\Delta t}\int
d\baeta_{1i}\, d\eta_{1i} \enskip e^{-\baeta_{1i} \eta_{1i} \Delta t} \nl
&\qquad \times  \exp\big[-i\, e^{i\pi/4} (\bpsi_{i+1}\eta_{1i} + \bar{\eta}_{1i}\theta_i) \Delta t \big], \nl  %\label{eqn:hs-1}
e^{-i \bar{\theta}_{i+1}\psi_{i} \Delta t}  &=
\frac{1}{\Delta t}\int d\baeta_{2i}\, d\eta_{2i} \enskip e^{-\baeta_{2i} \eta_{2i} \Delta t} \nl
&\qquad \times  \exp\big[-i \,e^{i\pi/4} (\bar{\theta}_{i+1}\eta_{2i} + \bar{\eta}_{2i}\psi_i\,)\Delta t\big],
\label{eqn:hs-2}
\end{align}
where $\{\eta_{1i},\,\baeta_{1i}, \, \eta_{2i},\,\baeta_{2i}\}$ are time-dependent
auxiliary Grassmann fields (AGFs) for $0 \leq i < N$.
Equation~\eqref{eqn:hs-2} can be regarded as a fermionic version of
the Hubbard-Stratonovich transformation for $\mathcal{U}_{_{\rm SB}}$.\cite{Sha045053}
Inserting \Eq{eqn:hs-2} into \Eq{pro:int}, we have
\begin{align} \label{pro:sd}
\mathcal{U}_{_{\rm SB}} (t_{i+1}, t_i) &=
\frac{1}{(\Delta t)^2}\int d\baeta_{1i} d\eta_{1i}d\baeta_{2i} d\eta_{2i}  \, e^{-\baeta_{1i} \eta_{1i} \Delta t} \nl
&\qquad  \times  e^{-\baeta_{2i} \eta_{2i} \Delta t} \, e^{-i e^{i\pi/4} (\bpsi_{i+1}\eta_{1i} + \baeta_{2i} \psi_i)\Delta t} \nl
&\qquad  \times  e^{-i e^{i\pi/4} (\btheta_{i+1} \eta_{2i} + \baeta_{1i}\theta_i) \Delta t}.
\end{align}
In the continuous time limit, the forward propagator of \Eq{pro:total} becomes
a weighted average over the AGFs
$\{\bm{\eta}, \bm\baeta \} = \{\eta_{j\tau}, \baeta_{j\tau} \}$.
\begin{align} \label{pro:con}
\mathcal{U}_{_{\rm T}}(t,\, 0) &= \int \mathcal{D} \,\bm\baeta \, \mathcal{D} \,\bm\eta \,
e^{-\int_{0}^{t} \bm \baeta_\tau \bm \eta_\tau d\tau } \,
\tilde{{U}}_{_{\rm S}}^f(t,\,0) \, \tilde{{U}}_{_{\rm B}}^f(t,\,0) \nonumber \\
&\equiv \la \, \tilde{{U}}_{_{\rm S}}^f(t,\,0) \, \tilde{{U}}_{_{\rm B}}^f(t,\,0) \, \ra.
\end{align}
Here, $\la \ldots \ra$ denotes the stochastic average over the AGFs.
With \Eq{pro:con}, the system and bath are formally decoupled from each other.
Instead, they are coupled to the AGFs $\{\eta_{j\tau},\,\baeta_{j\tau}\}\, (j=1,\,2)$,
where $\tilde{{U}}_{_{\rm S}}^f(t,\,0)$ and $\tilde{{U}}_{_{\rm B}}^f(t,\,0)$ are
the stochastic system and bath  propagators, respectively.
The weight
$e^{-\int_{0}^{t} \bm \baeta_\tau \bm \eta_\tau d\tau} =
e^{-\int_{0}^{t}\,(\baeta_{1\tau}\eta_{1\tau} + \baeta_{2\tau}\eta_{2\tau}) \,d\tau }$
is a Gaussian distribution.
The AGFs anti-commute with each other, i.e., $\eta_{j\tau}\eta_{j'\tau'} = - \eta_{j'\tau'}\eta_{j\tau}$;
and their stochastic averages satisfy the relations
$\la \baeta_{j\tau} \ra = \la\eta_{j\tau} \ra=0$ and $\la \eta_{jt}\baeta_{j'\tau} \ra = \delta_{jj'}\,\delta(t-\tau)$.
The backward propagator can be decoupled similarly by introducing the AGFs
$\{\eta_{j\tau}, \baeta_{j\tau} \}$ with $j=3,4$.

Suppose the initially density matrix has a factorized form of
$\rho_{_{\rm T}} (0) = {\rho}_{_{\rm S}} (0){\rho}_{_{\rm B}} (0)$,
the dynamics of the total system is given by,
\be \label{de-tot}
\rho_{_{\rm T}}(t)=\la \rho_{_{\rm S}} (t) \rho_{_{\rm B}} (t)\ra.
\ee
Here, $\la \cdots \ra$ denotes the stochastic average over the AGFs $\{\eta_{j\tau}, \baeta_{j\tau} \}$ ($j=1,\ldots,4$).
The time evolution of the decoupled stochastic system and bath density matrices is formally described by
\begin{align}
 \rho_{_{\rm S}} (t) &= \hat{\tilde{U}}_{_{\rm S}}^f(t,\,0)\rho_{_{\rm S}} (0)\,\hat{\tilde{U}}_{_{\rm S}}^b (0, t), \nl % \label{def:sys-pro} \nl
 \rho_{_{\rm B}} (t) &= \hat{\tilde{U}}_{_{\rm B}}^f(t,\,0)\rho_{_{\rm B}} (0)\,\hat{\tilde{U}}_{_{\rm B}}^b (0, t), \label{def:bat-pro}
\end{align}
where $\hat{\tilde{U}}_{_{\rm S/B}}^f$ and $\hat{\tilde{U}}_{_{\rm S/B}}^b$ are
the effective forward and backward evolution operators for the decoupled system/bath, respectively.
Specifically, the forward evolution operators take the time-ordered form of
\begin{align}
\hat{\tilde{U}}_{_{\rm S}}^{f} (t,\,0) &= \exp_{+} \Big\{-i\,\int_{0}^{t} d\tau \, \big[ H_{_{\rm S}} + e^{i\pi/4} \, \hlambda \nl
&\qquad \qquad  \times \left(\hc^\dag \eta_{_{1\tau}} + \bar{\eta}_{_{2\tau}} \hc \right) \big]  \Big\}, \label{pro:sf-1} \\
\hat{\tilde{U}}_{_{\rm B}}^f (t,\,0) &=  \exp_{+} \Big\{-i\,\int_{0}^{t} d\tau\, \big[ H_{_{\rm B}} + e^{i\pi/4}\,\mhlambda \nl
&\qquad \qquad \times \big(\bar{\eta}_{_{1\tau}}\hF + \hF^\dag {\eta}_{_{2\tau}} \big) \big] \Big\}. \label{pro:sf-2}
\end{align}
$\hat{\tilde{U}}_{_{\rm S/B}}^b(0,t)$ can be expressed similarly.
These operators result in the following SEOM for the decoupled stochastic system and bath density matrices:
\begin{align}
  \dot{\rho}_{_{\rm S}}
&= -i[\Hs, \rhos] + e^{-i\pi/4}
  \hlambda \, \big(\hc^\dag \eta_{1t} + \baeta_{2t} \hc \big)\rhos
\nl&\quad\,
  + e^{i\pi/4} \hlambda \, \rhos \big(\hc^\dag \eta_{3t} + \baeta_{4t} \hc \big),
\label{eom-rhos} \\
%%%%
  \dot{\rho}_{_{\rm B}}
&= -i[\Hb, \rhob]
   + e^{-i\pi/4} \mhlambda\,\big(\baeta_{1t} \hF + \hF^\dag \eta_{2t} \big)\rhob
\nl&\quad\,
  + e^{i\pi/4}\mhlambda\, \rhob
  \big(\baeta_{3t} \hF + \hF^\dag \eta_{4t} \big).
  \label{eom-rhob}
\end{align}
The system reduced density matrix is obtained as
\be \label{rdm}
\barrhos= \la \rho_{_{\rm S}} \trb(\rho_{_{\rm B}})\ra.
\ee
Obviously, $\trb(\rho_{_{\rm B}})$ is the weight of each quantum trajectory of $\rhos$.
Therefore, although there is no explicit coupling between system and bath in \Eqs{eom-rhos} and \eqref{eom-rhob},
the evolution of the fermion bath still affects the system dynamics
through the stochastic average of \Eq{rdm}.
Define $\trhos \equiv \rho_{_{\rm S}} \trb(\rho_{_{\rm B}})$,
so that $\brhos = \la \trhos \ra$, i.e.,
the quantum trajectories of $\trhos$ are equally weighted.

To verify the exactness of the stochastic decoupling,
we show the original \Sch equation of \Eq{def:total-eom} can be precisely
recovered by \Eqs{eom-rhos} and \eqref{eom-rhob}.
Consider the Ito's formula:%%%
\cite{Klo92}
%%%Xiao: cite ref for Ito's formula
%
\be
  d (\rhos \rhob) = d\rhos \rhob + \rhos d\rhob + d\rhos d\rhob. \label{pro-ito}
\ee
where $d\rhos d\rhob \sim \mathcal{O}(dt)$, and thus cannot be neglected.
Taking the average over all the AGFs on both sides of \Eq{pro-ito}, we have
\begin{align} \label{pr-1}
d\rho_{_{\rm T}} %&= \la d\rhos \rhob \ra + \la \rhos d\rhob \ra+ \la d\rhos d\rhob \ra \nl
&= -i\left[H_{_{\rm S}} + H_{_{\rm B}},\, \rho_{_{\rm T}}\right] dt \nl
&\quad-i\big\la \big(\hc^\dag  \eta_1 + \baeta_{2}\hc \big) \big(\baeta_{1} \hF + \hF^\dag \eta_{2} \big) \big \ra  \, \la \rhos \rhob \ra (dt)^2\nl
&\quad+i\la \rhos \rhob \, \ra \big \la
 \big(\hc^\dag \eta_{3} + \baeta_{4}\hc \big)\big(\baeta_{3} \hF + \hF^\dag \eta_{4} \big) \big \ra (dt)^2\nl
 &= -i \left[ H_{_{\rm S}} + H_{_{\rm B}} + H_{_{\rm SB}}, \, \rho_{_{\rm T}}\right] dt.
\end{align}
Here, we have used the causality principle that $\rhos(t)$ and $\rhob(t)$
depend only on $\{\baeta_{j\tau},\eta_{j\tau}\}$ at $\tau<t$,
as well as the equalities
$\rhos (\baeta_{1} \hF + \hF^\dag \eta_{2}) = (\baeta_{1} \hF + \hF^\dag \eta_{2}) \rhos$ and
$\rhob (\hc^\dag  \eta_{3t} + \baeta_{4t} \hc) = (\hc^\dag  \eta_{3t} + \baeta_{4t} \hc) \rhob$.
These equalities hold true because $\rhos$ ($\rhob$) and the operators
$\hF$ and $\hF^\dag$ ($\hc$ and $\hc^\dag$) belong to different physical spaces,
and the product of an AGF and a creation/annihilation operator
commutes with any function of AGFs.

In relation to the three contributions on the right-hand side of \Eq{pro-ito},
$\la d\rhos \rhob \ra$ and $ \la \rhos d\rhob \ra$
result in the dynamics due to the pure system and pure bath,
$[H_{_{\rm S}} + H_{_{\rm B}},\, \rho_{_{\rm T}}]$;
while $\la d\rhos d\rhob \ra$ gives rise to the dynamics
due to the system-bath interaction, $[H_{_{\rm SB}},\, \rho_{_{\rm T}}]$.

\subsection{Capturing the non-Markovian memory of fermionic environment}

In practice, the bath is considered to have an infinite DOF,
and thus we normally avoid solving \Eq{eom-rhob} explicitly.
In \Eq{rdm}, $\trb(\rhob)$ is the weight of the trajectory of $\rhos$,
which includes the AGFs $\{\eta_{j\tau},\,\baeta_{j\tau}\}$.
It captures the non-Markovian memory of the fermion bath on the reduced system dynamics,
and it can be evaluated by analyzing \Eq{eom-rhob}.

In the $\Hb$-interaction picture, the formal solution of \Eq{eom-rhob} is worked out as
\be \label{def:irhob}
\rhob(t) = \huz(t)\,\htb^f(t,0)\,\rhob(0)\,\htb^b(0,t)\,\hat{U}_{_{\rm B}}^{(0)\dag}(t).
\ee
Here, $\huz(t) = \exp(-i\Hb t)$ is the evolution operator for the isolated bath,
with $\hF(t) \equiv \huz(t)\hF\,\hat{U}_{_{\rm B}}^{(0)\dag}(t)$.
Using the cyclic permutation invariance of trace operation, we have
\be \label{def:trace}
\trb(\rhob)=\trb \Big[\,\hat{\tilde{U}}_{_{\rm B}}^f(t,0)\, \rhob(0)\, \hat{\tilde{U}}_{_{\rm B}}^b(0,t)\,\Big].
\ee
Initially, the isolated bath is assumed to be in a thermal equilibrium state,
i.e., $\rhob(0)= \rhob^{\rm eq}= e^{-\beta (\Hb-\mu^{\rm eq}_{_{\rm B}} \hat{N}_{_{\rm B}})}/Z$,
where $\hat{N}_{_{\rm B}}$ is the particle number operator,
$\mu^{\rm eq}_{_{\rm B}}$ is the equilibrium chemical potential,
$\beta=1/T$ is the inverse temperature,
and $Z = \trb[e^{-\beta (\Hb-\mu^{\rm eq}_{_{\rm B}} \hat{N}_{_{\rm B}})}]$
is the partition function of the bath, respectively.

For a non-interacting fermion bath which satisfies the Gaussian statistics,
\Eq{eom-rhob} can be formally solved by using the Magnus expansion\cite{Tan06}
and the Baker-Campbell-Hausdorff formula.\cite{Gre96}
This leads to the following expression of $\trb(\rhob)$:
\be \label{tra-rhob}
\trb (\rhob) = e^{\int_{0}^{t} d\tau [(\baeta_{1\tau}-i\baeta_{3\tau})g^-_\tau +
(\eta_{2\tau} - i\eta_{4\tau}) g^+_\tau ] }.
\ee
The non-Markovian memory effects are accounted for by the following memory-convoluted AGFs:\cite{Han19050601}
\be\label{def-gt}
\begin{split}
  g^-_t & = \lambda^{-1} \int_{0}^t
 \left\{ [\Cp(t-\tau)]^\ast \eta_{4\tau}
 -i \Cm (t-\tau) \,\eta_{2\tau} \!  \right\} d\tau,
\\
  g^+_t & = \lambda^{-1}\int_{0}^t
 \left\{ [\Cm(t-\tau)]^\ast \baeta_{3\tau}
 -i \Cp (t-\tau) \,\baeta_{1\tau} \! \right\} d\tau,
\end{split}
\ee
where $\Cp(t-\tau) = \trb \big[\hF^\dag(t)\, \hF(\tau)\, \rhob^{\rm eq}\big]$ and
$\Cm(t-\tau) = \trb \big[\hF(t)\, \hF^\dag(\tau)\, \rhob^{\rm eq}\big]$
are two-time correlation functions of the fermion bath.
They are related to the bath spectral density,
$J(\omega) \equiv \pi \sum_{k} |t_k|^2 \delta (\omega-\epsilon_k)$,
via the fluctuation-dissipation theorem as follows,
\be \label{fdt-1}
C^\sigma(t) = \int_{-\infty}^{+\infty} d \omega \, e^{\sigma i \omega t}f^\sigma(\omega) J(\omega).
\ee
Here, $\sigma=\pm$ and
\be \label{def:fermi-f}
f^\sigma(\omega) = \frac{1}{1+e^{\sigma \beta (\omega-\mu)}},
\ee
is the Fermi function for electron $(\sigma=+)$ or hole $(\sigma=-)$
at temperature $T = 1/\beta$, and $\mu$ is the bath chemical potential.
%
%All environmental
%influences are encoded in \Eq{tra-rhob}, which is directly related
%to the influential functional of the fermionic bath \cite{Jin07134113}.
%
%To avoid the direct evaluation of $\trb[\rhob(t)]$ by \Eq{eom-rhob}, we apply the
%Girsanov transformation to adsorb $\trb[\rhob(t)]$ into the probability distribution
%of Grassmann Gaussian processes.

To obtain quantum dynamical trajectories of the reduced system with equal weights,
we refer to the SEOM for bosonic environments.\cite{Sha045053,Yan161}
The Girsanov transformation has been utilized to absorb the weight function $\trb(\rhob)$
into the dynamical variable.\cite{Gat912152,Sha045053}
The fermionic analogue is expressed as
\begin{align} \label{def:Girsanov}
\la\, e^{-\bar{\theta}\eta}f(\theta,\,\bar{\theta})\, \ra
%e^{-\bar{\theta}\eta}f(\theta,\,\bar{\theta})\nl
&= \int d\bar{\theta}\, d\theta \, e^{-\bar{\theta}(\theta+\eta)}f(\theta,\,\bar{\theta}) \nl
&= \la f(\theta-\eta,\,\bar{\theta}) \ra,
\end{align}
where $f(\theta,\,\bar{\theta})$ is any analytic function of the g-numbers
$\theta$ and $\bar{\theta}$, and $\eta$ is another g-number.
Applying \Eq{def:Girsanov} to the system reduced density matrix of \Eq{rdm},
and making use of \Eq{tra-rhob}, we have
\begin{align}
  \brhos (t) &= \la \,\rhos(t)\, \trb(\rho_{_{\rm B}}) \, \ra \nl
   &= \int \mathcal{D} {\bm \baeta}\,\mathcal{D} {\bm \eta}
   \,\, e^{-\int_{0}^t \bm{\baeta}_{\tau} \bm{\eta}_{\tau} {\rm d}\tau} \, \rhos(\bm\eta, \bm \baeta)\, \trb(\rho_{_{\rm B}})  \nl
   &= \int \mathcal{D} {\bm \baeta'}\,\mathcal{D} {\bm \eta'}
   \,\, e^{-\int_{0}^t \bm{\baeta'}_{\tau} \bm{\eta'}_{\tau} {\rm d}\tau} \, \trhos(\bm\eta', \bm \baeta') \nl
  &= \la \trhos \ra\,.  \label{aver-gir}
\end{align}
Here, the third equality involves the following transformation of variables:
\begin{align}
\eta_{1\tau}' & = \eta_{1\tau} - \gm_\tau \nl
\baeta_{2\tau}' & = \baeta_{2\tau} +\gp_\tau \nl
\eta_{3\tau}' & = \eta_{3\tau} + i\gm_\tau \nl
\baeta_{4\tau}'& = \baeta_{4\tau} - i\gp_\tau.
\end{align}
Consequently, the dissipative dynamics of the system is described by the following SEOM
\begin{align}
 \dot{{\trho}}_{_{\rm S}}
&= -i[ \Hs, \trhos]
   + e^{-i\pi/4}\, \hlambda\big\{\hc^\dag \gm_t - \gp_t \hc, \,\trhos \big\}
\nl & \quad
+ e^{-i\pi/4}\, \hlambda \big(\hc^\dag \eta_{1t} + \baeta_{2t} \hc \big)\,\trhos  \nl
&\quad + e^{i\pi/4}\, \hlambda\, \trhos  \big(\hc^\dag \eta_{3t} + \baeta_{4t} \hc \big).
\label{eom-trhos}
\end{align}
The derivation is formally analogous to that for a boson bath,
\emph{e.g.}, Eq.~(17) in Ref.~\onlinecite{Sha045053}.
The non-Markovian bath memory is captured by the memory-convoluted AGFs $\{g_t^{\pm}\}$,
and the instantaneous AGFs $\{\eta_{jt}, \baeta_{jt}\}$
characterize the fluctuations exerted by the bath to the system.

As will be verified in the next subsection, \Eq{eom-trhos} is formally exact,
as long as \Eq{tra-rhob} is valid, i.e., if the fermion bath satisfies
the Gaussian statistics and the quantum trajectories of $\trhos$ are equally weighted.
However, the direct numerical implementation of \Eq{eom-trhos} still faces fundamental difficulties,
because of the problem in the realization of AGFs.
Such difficulties severely hinder the practical use of \Eq{eom-trhos}
or its analogues for open fermionic systems.\cite{Che13052108,Hsi18014103}
Consequently, all the previous efforts on fermionic SEOM ended up with formal derivations,\cite{Zha12032116, Lin17180401}
or transformation to deterministic approaches.\cite{Sue151408,Hsi18014103, Hsi18014104}

\subsection{Equivalence between the formally exact SEOM and HEOM formalisms} \label{seom-heom}

We now analytically verify the equivalence between the SEOM of \Eq{eom-trhos}
and the rigorous HEOM. Similar to the derivation for a boson bath,\cite{Sch151, Yan161}
we take the average on both sides of \Eq{eom-trhos} over all the AGFs $\{\eta_{j\tau},\,\baeta_{j\tau}\}$.
The reduced system dynamics is determined by the following EOM:
\begin{align}
\la{\,\dot{\tilde{\rho}}_{_{\rm S}} }\ra
 %&= -i[ \Hs, \la \trhos \ra]
 %  + e^{-i\pi/4} \hlambda \la \big\{\hc^\dag \gm - \gp \hc, \,\trhos \big\} \ra \nl
   &= -i[ \Hs, \la \trhos \ra]
   + e^{-i\pi/4}\hlambda \Big(\hc^\dag\la \gm_t \trhos \ra + \la \trhos \gm_t \ra\, \hc^\dag \nl
   &\quad -\hc\la \gp_t \trhos \ra - \la \trhos \gp_t \ra \, \hc\, \Big)\, .
   \label{eom-av}
\end{align}
where $\la \baeta_{jt}\,\trhos \ra = \la \eta_{jt}\, \trhos\ra  = 0$ because of the causality relation.
Such an EOM is not self-closed, because $\la \trhos g^+_t \ra $ and $\la g^-_t \trhos \ra$
are not an explicit function of $\la \trhos \ra$.
To proceed, we unravel the bath correlation function $C^{\sigma} (t)$ by
a number of exponential functions,
\be
C^\sigma (t) = \sum\limits_{m=1}^M C_m^\sigma(t) =\sum\limits_{m=1}^M A_m^\sigma e^{\gamma_m^\sigma t},
\label{cexp}
\ee
with the symmetry $\gamma^+_m = (\gamma^-_m)^*$ satisfied,
and the memory-convoluted AGFs are decomposed accordingly as $g^\sigma_t = \sum_{m=1}^{M} g_m^\sigma (t)$.
Each of the components, ${g}^\sigma_m\,(t)$, satisfies a self-closed EOM
\begin{align} \label{mem}
  \dot{g}^-_m &= \lambda^{-1}\left[ -i A_m^{-}\, \eta_{2t} + (A_m^{+})^\ast \, \eta_{4t} \right]
   + \gamma_m^{-}\, g^-_m, \nl
  \dot{g}^+_m &= \lambda^{-1}\left[ -i A_m^{+}\, \bar{\eta}_{1t} + (A_m^{-})^\ast\, \bar{\eta}_{3t} \right]
   + \gamma_m^{+}\, g^+_m.
\end{align}

In the path-integral formulation of HEOM,
an $(I+J)$th-tier auxiliary density operator (ADO) is defined by
($I$ and $J$ are arbitrary non-negative integers):
\begin{align}
\rho_{m_1 \ldots m_I n_1 \cdots n_J}^{(-\cdots-+\cdots+)} &\equiv
\int \mD \bm\bpsi \, \mD\bm\psi\, \mD \bm\bpsi' \, \mD\bm\psi'\,
e^{iS_{\!f}} \mathcal{F}_{_{\rm FV}} e^{-iS_{\rm b}} \nl
&\qquad \times \mathcal{B}^{-}_{m_I} \cdots \mathcal{B}^{-}_{m_1}
\mathcal{B}^{+}_{n_J} \cdots \mathcal{B}^{+}_{n_1}\, \rhos(0).
\label{def:ado-1}
\end{align}
Here, $\{{\bm \bpsi},\,{\bm \bpsi}'\}\equiv\{{ \bpsi}_\tau,\,{\bpsi}'_\tau\}$ and
$\{{\bm \psi},\,{\bm \psi}'\}\equiv\{{ \psi}_\tau,\,{\psi}'_\tau\}$ are g-numbers
associated with $\hc^\dag$ and $\hc$, respectively;
$S_{\!f}$ and $S_b$ are the forward and backward
action functionals associated with $\Hs$,
$\mathcal{F}_{_{\rm FV}}$ is the Feynman-Vernon influence functional,\cite{Fey63118}
and $\mathcal{B}^{-}_m$ and $\mathcal{B}^{+}_n$ are the generating functionals:
\begin{align}
\mathcal{B}^{-}_m &=  -i \int_{0}^t d\tau  \left[ A^-_m  \, \psi_\tau - \big(A^{+}_m \big)^\ast  \, \psi'_\tau
\right]  e^{\gamma_m^{-}(t-\tau)}, \nl
\mathcal{B}^{+}_n &=  -i \int_{0}^t d\tau  \left[ A^+_n  \, \bpsi_\tau - \big(A^{-}_n \big)^\ast \, \bpsi'_\tau
\right]  e^{\gamma_n^{+}(t-\tau)}.  \label{def-b-terms-1}
\end{align}

In the stochastic framework, such an $(I+J)$th-tier ADO is retrieved as follows:
\begin{align}
 \rho_{m_1 \ldots m_I n_1 \cdots n_J}^{(-\cdots-+\cdots+)} &=
(e^{i\pi/4}\,\hlambda)^{I+J}\, \la \gm_{m_1} \cdots \gm_{m_I} \trhos \,\gp_{n_1} \cdots \gp_{n_J} \ra \nl
&= \la \, \Gm\, \trhos \, \Gp \ra \,.
\label{ado-seom-1}
\end{align}
%
%Here, $\{m_i\}$ and $\{n_j\}$ respectively correspond
%to different modes of reservoir; and $g_m^\sigma(t)$ are considered as the analogue of
%$\mathcal{B}_m^\sigma(t)$. For the convenience of computation, we
For brevity, we will use the notation $\tilde{\prod}$ to represent the ordered products,
$\Gm \equiv (e^{i\pi/4}\hlambda)^{I}\, \tilde{\prod}_{i=1}^{I} \gm_{m_i}$
and $\Gp \equiv (e^{i\pi/4}\hlambda)^{J}\, \tilde{\prod}_{j=1}^{J} \gp_{n_j}$.
In the ordered products, the indices $\{m_i\}$ and $\{n_j\}$ are arranged in
the ascending order of $i$ and $j$. Accordingly, \Eq{eom-av} is rewritten as
\begin{align}
\dot{\rho}_{_{\rm S}}^{(0)}
   &= -i[ \Hs, \rhos^{(0)}]
   -i \sum\limits_{m=1} \Big(\hc^\dag \rho_m^{(-)}-\rho_m^{(-)} \hc^\dag \nl
   &\qquad \qquad \qquad \qquad  \qquad
   + \hc\rho_m^{(+)} - \rho_m^{(+)} \hc\,\Big), \label{ADM-0rd}
\end{align}
where $\rhos^{(0)} = \la \trhos \ra = \brhos$ is the reduced density matrix
of the system, and $\rho_m^{(-)} = e^{i\pi/4}\hlambda\la g_m^{-} \trhos \ra$
and $\rho_m^{(+)} = e^{i\pi/4}\hlambda\la  \trhos g_m^{+}\ra$ are first-tier ADOs, respectively.
Besides, an important equality is utilized in the derivation,
\be \label{av_com}
\la f(t) g_m^{\sigma}\ra = -\la g_m^{\sigma} f(t) \ra,
\ee
which holds for any analytic function $f(t)$ of the AGFs $\{\eta_{j\tau},\,\baeta_{j\tau}\}\,(j=1,\cdots,4)$.

To derive the EOM for the first-tier ADOs, we use the Ito's formula
to explore the differential of $\la \trhos g^\sigma_m \ra$, which yields
\be \label{t-deriv}
d\la\trhos g^\sigma_m\ra
= \la d\trhos\, g^\sigma_m \ra+  \la \trhos \, dg^\sigma_m\ra +  \la d \trhos \, d g^\sigma_m \ra.
\ee 	
Here, the first term on the right gives rise to ADOs at the second tier,
while the last term retrieves the zeroth-tier ADO.
Without loss of generality, the differential of
$\rho_{m_1 \ldots m_I n_1 \cdots n_J}^{(-\cdots-+\cdots+)}$
can be expressed as a sum of three parts,
\begin{align}
  d \rho_{m_1 \ldots m_I n_1 \cdots n_J}^{(-\cdots-+\cdots+)} & =
  d\la \, \Gm \trhos \, \Gp \ra \nl
  &= \big[ \la \Gm\, d\trhos \, \Gp \ra  + \la d\Gm\, \trhos \Gp \ra + \la \Gm\, \trhos  d\Gp\ra \nl
  & \quad +\la d\Gm\, d\trhos \, \Gp \ra + \la \Gm\, d\trhos \, d\Gp \ra\big] \nl
  &= \Xi_1 + \Xi_2 + \Xi_3.
  \label{heom-ado-1}
\end{align}
Here, $\Xi_1$, $\Xi_2$ and $\Xi_3$ collect the contributions of ADOs at the $(I+J + 1)$th,
$(I+J)$th and $(I+J - 1)$th tiers, respectively;
whereas $\la d\Gm\, \trhos d\Gp \ra \sim \mathcal{O}(dt^2)$ makes zero contribution.
Presuming that the AGFs commute with the system creation and annihilation operators
($\hat{c}^\dag$ and $\hat{c}$) in the SEOM of \Eq{eom-trhos}, we have
\begin{align}
 \Xi_1 &=\la \Gm d\trhos \,\Gp \ra \nl
   &= -i \big[ \Hs,  \rho_{m_1 \ldots m_I n_1 \cdots n_J}^{(-\cdots-+\cdots+)} \big] dt
    -i \sum_{r=1}^M \, \Big[ \hc^\dag \, \rho_{m_1 \ldots m_I r n_1 \cdots n_J}^{(-\cdots--+\cdots+)}\nl
   &\quad
  -(-1)^{I+J} \rho_{m_1 \ldots m_I r n_1 \cdots n_J}^{(-\cdots--+\cdots+)} \hc^\dag
  + (-1)^{I+J}  \hc\, \rho_{m_1 \ldots m_I r n_1 \cdots n_J}^{(-\cdots-++\cdots+)}\nl
 &\quad -\rho_{m_1 \ldots m_I r n_1 \cdots n_J}^{(-\cdots-++\cdots+)} \,\hc \, \Big] dt.
  \label{term-1}
\end{align}
Define the ordered products
${\mathcal{G}}^{-\,\succ}_{\bm{m}_i}\equiv
(e^{i\pi/4}\hlambda)^{i} \tilde{\prod}_{i'=1}^{i} g^-_{m_{i'}}$ and
${\mathcal{G}}^{-\,\prec}_{\bm{m}_i} \equiv
(e^{i\pi/4}\hlambda)^{I-i+1} \tilde{\prod}_{i'=i}^{I} g^-_{m_{i'}}$,
with which we have
$\Gm = {\mathcal{G}}^{-\succ}_{\bm{m}_{i-1}}\gm_{m_{i}}{\mathcal{G}}^{-\prec}_{\bm{m}_{i+1}}$.
Similarly, ${\mathcal{G}}^{+\succ}_{\bm{n}_j}$ and ${\mathcal{G}}^{+\prec}_{\bm{n}_j}$
are introduced, and $\Gp = {\mathcal{G}}^{+\succ}_{\bm{n}_{j-1}}\gp_{n_j}{\mathcal{G}}^{+\prec}_{\bm{n}_{j+1}}$.
Consequently, $\Xi_2$ and $\Xi_3$ are simplified as
\begin{align}
  \Xi_2 &= \la d\Gm\, \trhos \, \Gp \ra +\la \Gm\, \trhos \, d\Gp \ra \nl
  &= \sum_{i=1}^I \Big\la {\mathcal{G}}^{-\succ}_{\bm{m}_{i-1}} d\gm_{m_i} {\mathcal{G}}^{-\prec}_{\bm{m}_{i+1}}\trhos \Gp \Big\ra \nl
  & \quad + \sum_{j=1}^J \Big\la \Gm \trhos {\mathcal{G}}^{+\succ}_{\bm{n}_{j-1}}  d\gp_{n_j} {\mathcal{G}}^{+\prec}_{\bm{n}_{j+1}} \Big\ra \nl
  &=  \Big( \sum_{i=1}^I \gamma_{m_i}^{-} + \sum_{j=1}^J \gamma_{n_j}^{+} \Big)
 \rho_{m_1 \ldots m_I n_1 \cdots n_J}^{(-\cdots-+\cdots+)} \, dt,
 \label{term-2}
\end{align}
and
\begin{align}
 \Xi_3  &=\la d\Gm\, d\trhos \, \Gp \ra + \la \Gm\, d\trhos \, d\Gp \ra \nl
  &= -i\sum_{i=1}^I  \,
  \Big[ A_{m_i}^{-}
  \hc \, \big\la {\mathcal{G}}^{-\succ}_{\bm{m}_{i-1}} \eta_{2t}\, {\mathcal{G}}^{-\prec}_{\bm{m}_{i+1}}\,\bar{\eta}_{2t} \,\trhos \,\Gp \big\ra \nl
 %&\quad
 &\quad- (A^{+}_{m_i})^\ast \big\la {\mathcal{G}}^{-\succ}_{\bm{m}_{i-1}}\, \eta_{4t}\, {\mathcal{G}}^{-\prec}_{\bm{m}_{i+1}}
 \trhos\, \bar{\eta}_{4t}\, \Gp \big\ra \, \hc \,\Big](dt)^2 \nl
 &\quad -i\,\sum_{j=1}^J  \Big[ A_{n_j}^{+} \hc^\dag
 \, \big\la \Gm \,\eta_{1t}\, \trhos\, {\mathcal{G}}^{-\succ}_{\bm{n}_{j-1}}\bar{\eta}_{1t} \gp_{n_{j}}
 {\mathcal{G}}^{+\prec}_{\bm{n}_{j+1}} \big\ra \nl
 %&\quad
 &\quad - (A^{-}_{n_j})^\ast \big\la \Gm \trhos\, \eta_{3t}
 \,{\mathcal{G}}^{+\succ}_{\bm{n}_{j-1}} \,\bar{\eta}_{3t}\,
 {\mathcal{G}}^{+\prec}_{\bm{n}_{j+1}} \big\ra \, \hc^\dag \,\Big](dt)^2\nl
  &= -i \sum_{i=1}^I \Big[ A_{m_i}^{-} (-1)^{I-i} \, \hc \,
 \rho_{m_1\ldots m_{i-1} m_{i+1}\cdots n_J}^{(-\cdots--\cdots+)} \nl
 &\quad- (A^{+}_{m_i})^\ast (-1)^{i-1+J}\, \rho_{m_1\ldots m_{i-1} m_{i+1}\cdots n_J}^{(-\cdots--\cdots+)} \, \hc\, \Big] dt \nl
 &\quad -i \sum_{j=1}^J \Big[ A_{n_j}^{+}\, (-1)^{I+J-j} \, \hc^\dag \,
 \rho_{m_1\ldots n_{j-1} n_{j+1}\cdots n_J}^{(-\cdots++\cdots+)} \nl
&\quad- (A^{-}_{n_j})^\ast \, (-1)^{j-1}\,
 \rho_{m_1\ldots n_{j-1} n_{j+1}\cdots n_J}^{(-\cdots++\cdots+)} \, \hc^\dag\, \Big] dt.
 \label{term-3}
\end{align}
Here, we have utilized the causality relation to obtain an important equality,
\begin{align}
& \big\la {\mathcal{G}}^{-\succ}_{\bm{m}_{i-1}}  \eta_{2t} {\mathcal{G}}^{-\prec}_{\bm{m}_{i+1}}  \bar{\eta}_{2t}\, \trhos\, \Gp \big\ra (dt)^2\nl
 = \enskip &(-1)^{I-i} \, \la \eta_{2t} \, \baeta_{2t}\ra \,
 \big\la {\mathcal{G}}^{-\succ}_{\bm{m}_{i-1}}\,{\mathcal{G}}^{-\prec}_{\bm{m}_{i+1}}\trhos \, \Gp \big\ra (dt)^2  \nl
 =\enskip  &(-1)^{I-i}  \, \rho_{m_1\ldots m_{i-1} m_{i+1}\cdots n_J}^{(-\cdots--\cdots+)}\, dt.
\end{align}
Equation~\eqref{heom-ado-1} is finally recast into a compact form of
\begin{align}
 \dot{\rho}_{m_1 \ldots m_I n_1 \cdots n_J}^{(-\cdots-+\cdots+)} &=\Big(-i \mathcal{L}_{_{\rm S}} + \sum_{i=1}^I \gamma_{m_i}^-
 + \sum_{j=1}^J \gamma_{n_j}^+ \Big)\rho_{m_1 \ldots m_I n_1 \cdots n_J}^{(-\cdots-+\cdots+)} \nl
 &\quad+ \sum_{i=1}^I  \mathcal{C}^-_{m_i}\, \rho_{m_1 \ldots m_{i-1} m_{i+1} \cdots n_J}^{(-\cdots--\cdots+)} \nl
 &\quad+ \sum_{j=1}^J  \mathcal{C}^+_{n_j}\, \rho_{m_1 \ldots n_{j-1} n_{j+1} \cdots n_J}^{(-\cdots++\cdots+)} \nl
 &\quad+ \sum_{\sigma=+,-} \sum_{r=1}^M \mathcal{A}^\sigma_r \, \rho_{m_1 \ldots m_I r n_1 \cdots n_J}^{(-\cdots-\sigma+\cdots+)}.
  \label{heom-1}
\end{align}
Here,  %our referring to the related concepts in standard HEOM,
$\mathcal{L}_{_{\rm S}} \star \equiv [\Hs,\, \star]$ is defined as
the system Liouvillian,
and the super-operators $\mathcal{A}_j^\sigma$ and $\mathcal{C}_{j}^\sigma$
are given by \Eq{term-1} and \Eq{term-3}, respectively.
Equation~\eqref{heom-1} exactly reproduces Eq.~(1) of Ref.~\onlinecite{Han18234108}.
This proves that the SEOM of \Eq{eom-trhos} is rigorous,
as it is in principle equivalent to the formally exact HEOM formalism for fermionic open systems.

\section{A numerically feasible fermionic SEOM method} \label{sec:auxiliary_space}

\subsection{A minimal auxiliary space mapping scheme for the AGFs}

In contrast with the bosonic case, the applications of the fermionic SEOM
has been prohibited because of the numerical difficulty in realization of g-numbers.
Different from the c-numbers, the g-numbers cannot be represented by scalars.
Instead, it would require the use of $N$ mutually anti-commutative matrices of
the size $2^N\times 2^N$ to represent a set of $N$ g-numbers.
The computer memory required to store these matrices will soon become too
large with the increase of $N$.

To enable the direct stochastic simulation of fermionic dissipative dynamics
by using the SEOM of \Eq{eom-trhos} or its analogues,
a mapping approach has been established in our previous work.\cite{Han19050601}
In this subsection, we elaborate on this approach by providing more details and deeper insights.

Intuitively, one would attempt to ``simplify'' the time-dependent AGFs $\{\eta_{j\tau},\bar{\eta}_{j\tau}\}$
by separating the time dependence from the Grassmann character via the following mapping:
\be \label{tran_1}
 \eta_{j\tau} \mapsto v_{j\tau} \xi_j, \enskip \ \ \baeta_{j\tau} \mapsto v_{j\tau} \bar{\xi}_j \quad (\, j=1,\cdots,\,4 \,),
\ee
where $\{{\xi}_{j},\bar{\xi}_{j}\}$ are time-independent g-numbers,
and $\{v_{j\tau}\}$ are Gaussian c-number noises with
$\mathcal{M}(v_{j\tau})=0$ and $\mathcal{M}(v_{jt}\,v_{j'\tau}) = \delta_{jj'}\delta(t-\tau)$.
Here, $\mathcal{M}(\cdots)$ denotes the average over c-number noises.
The mapping of \Eq{tran_1} preserves the Grassmann character
$\eta_{j\tau}\baeta_{j'\tau} = -\baeta_{j'\tau} \eta_{j\tau}$,
and the statistical properties $\la\eta_{j\tau}\ra = \la\baeta_{j\tau}\ra =0$
and $\la \eta_{jt} \baeta_{j'\tau} \ra = \delta_{jj'}\delta(t-\tau)$.
Meanwhile, such a mapping drastically reduces the memory cost for
storing the AGFs, as the 8 time-independent g-numbers $\{\xi_j, \bar{\xi}_j\}$
can be represented by 8 matrices of the size $2^8\times 2^8$.
The system reduced density matrix can be obtained by
\be \label{def:average-2}
\la \trhos \ra = \int d \bar{\bm \xi}\, d{\bm \xi} \, e^{-\bar{\bm \xi}\, \bm{\xi}} \,\mathcal{M}(\trhos),
\ee
where $\{{\bm{\xi}}, \bar{\bm{\xi}}\} \equiv \{{{\xi}_j}, \bar{{\xi}_j}\}\, (j=1,\cdots,4)$.

Unfortunately, the mapping of \Eq{tran_1} cannot rigorously recover the result of \Eq{eom-rhos}.
This is because some important properties of the original AGFs, such as
\begin{align}
\la \eta_{jt}\baeta_{jt}\eta_{j'\tau}\baeta_{j'\tau}\ra &=
\la \eta_{jt}\baeta_{jt} \ra \la \eta_{j'\tau}\baeta_{j'\tau}\ra -
\la \eta_{jt}\baeta_{j'\tau} \ra \la \eta_{j'\tau} \baeta_{jt}\ra \nl
&= [\delta(0)]^2 - \delta_{jj'} [\delta(t-\tau)]^2,   \label{AGF-4th-moment-1}
\end{align}
are not preserved. The major drawback of \Eq{tran_1} can be understood by
considering the prototypical equation
\be
\dot{y} = y \left[D(t)\,\eta_t
+ \int_{0}^t C(\tau)\,\baeta_\tau {\rm d}\tau \right], \label{eom-y-1}
\ee
where $\eta_t$ and $\baeta_\tau$ are time-dependent AGFs,
and $C(t)$ and $D(t)$ are c-number functions.
In Ref.~\onlinecite{Han19050601} it has been shown that the mapping of \Eq{tran_1}
fails to reproduce the exact solution of \Eq{eom-y-1}.

To fix the above problem, we introduce a reduction procedure as a complement to \Eq{tran_1}.
Such a procedure can be formally described by a linear operator $\hat{r}$,
which reduces any matching g-number pair (such as $\xi_j \bar{\xi}_j$) to $1$
at each time step, i.e.,
\be \label{red}
 \hat{r}(1) = 1, \, \hat{r}(\xi_{j}) = \xi_{j}, \,  \hat{r}(\bar{\xi}_j)= \bar{\xi}_j, \, \hat{r}(\xi_j\bar{\xi}_j) =1.
\ee
With the reduction procedure, any even-order moment of AGFs,
including the one on the left-hand side of \Eq{AGF-4th-moment-1},
is correctly reproduced.
By applying \Eq{red} at each time step, the solution of \Eq{eom-y-1}
is exactly recovered; see the Supplemental Material of Ref.~\onlinecite{Han19050601}.

With the mapping of \Eq{tran_1}, $\trhos$ in the SEOM of \Eq{eom-trhos}
becomes an analytic function of $\{\xi_j, \bar{\xi}_j\}$.
The reduction procedure can be described by introducing the operators
$\{\hat{\xi}_j, \hat{\bar{\xi}}_j\}$, which can act on $\trhos$ from left or right:
\begin{align}\label{def-red}
\hat{\xi}_j \, \trhos &= \hat{r}(\xi_j\, \trhos), \enskip
\trhos \, \hat{\xi}_j  = \hat{r}(\trhos\, \xi_j), \nl
\hat{\bar{\xi}}_j \,\trhos  &= \hat{r}(\bar{\xi}_j\, \trhos ),\enskip
\trhos \, \hat{\bar{\xi}}_j  = \hat{r}(\trhos \, \bar{\xi}_j).
\end{align}
More specifically, the left action results in
\bea
\label{define-a-hat-2} \hat{\xi}_j\, 1 = \xi_{j},\enskip \hat{\xi}_j\, \xi_j &=& 0, \enskip \hat{\xi}_j\, \bar{\xi}_j=1 \nl
\hat{\barxi}_j\, 1 = \barxi_{j},\enskip \hat{\barxi}_j\, \barxi_j &=& 0, \enskip \hat{\barxi}_j\, {\xi}_j=-1.
\eea
Similarly, for the right side action, we have
\bea \label{define-a-hat-3}
1\,\hat{\xi}_j  = \xi_{j},\enskip \xi_j\, \hat{\xi}_j &=& 0, \enskip \bar{\xi}_j\, \hat{\xi}_j=-1 \nl
1\, \hat{\barxi}_j  = \barxi_{j},\enskip \barxi_j \, \hat{\barxi}_j &=& 0, \enskip  {\xi}_j\, \hat{\barxi}_j=1.
\eea
With \Eq{tran_1} and the reduction operation, the time-dependent AGFs
$\{\eta_{j\tau}, \baeta_{j\tau}\}$ are mapped to c-number noises $\{v_{j\tau}\}$
and a set of three elements $\{\xi_j,1,\bar{\xi}_j\}$.
Such a set is isomorphic to a minimal auxiliary space (MAS),
$S_j = \{ -1_j, 0_j, 1_j\}$,  via the following one-to-one correspondence:
\be \label{def:corres}
1 \mapsto |0_j\ra,\enskip \xi_j \mapsto |\!-\! 1_j\ra,\enskip \bar{\xi}_j \mapsto |1_j\ra.
\ee
Here, $|0_j\ra$, $|\!-\! 1_j\ra$ and $|1_j\ra$ represent the three pseudo-Fock-states, i.e.,
the pseudo-vacuum, pseudo-hole and pseudo-particle states,
of the $j$th pseudo-level, respectively.
Accordingly, the operators $\{\hat{\xi}_j,\hat{\bar{\xi}}_j\}$ correspond to
the pseudo-operators $\{X_j^-, X_j^+\}$:
\be \label{def:map-2}
 \hat{\xi}_j \mapsto X_{j}^{-},\enskip \hat{\bar{\xi}}_j \mapsto X_{j}^{+}.
\ee
The AGFs $\{\eta_{j\tau},\baeta_{j\tau}\}$ are finally represented by
\be \label{sub-2}
\eta_{j\tau} \mapsto v_{j\tau} \Xd_j, \ \ \baeta_{j\tau} \mapsto v_{j\tau} \Xu_j\enskip (j=1,\cdots,4).
\ee
Equation~\eqref{sub-2} is termed as the MAS mapping.

It is important to emphasize that, unlike normal operators
which are associated with physical observables,
the pseudo-operators are mathematical tools to assist tracking
the time order of AGFs. For this purpose,
the pseudo-operators are allowed to act on both the left and right
sides of a pseudo-Fock-state (denoted by a ket), and the left
and right actions may lead to different results. This is distinctly
different from a normal operator.
Specifically,
the left actions of the pseudo-operators on the pseudo-Fock-states yield
\be
\begin{aligned}
 \Xd_j |1_j\ra &= |0_j\ra,  &\Xd_j |0_j\ra &= |\! -\! 1_j\ra,  &\Xd_j |\! -\! 1_j\ra &= 0, \\
 \Xu_j |1_j\ra &= 0,  &\Xu_j |0_j\ra &= |1_j\ra,  &\Xu_j |\! -\! 1_j\ra &= -|0_j\ra.
  \label{ladder-spin-1}
 \end{aligned}
\ee
The actions from the right lead to
\be
 \begin{aligned}
 |1_j\ra \Xd_j &= -| 0_j \ra,  &| 0_j \ra \Xd_j  & = |\! -\! 1_j\ra,  &|\! -\! 1_j\ra \Xd_j &= 0, \\
 |1_j \ra \Xu_j &= 0,  &| 0_j \ra \Xu_j  &= | 1_j \ra, &|\! -\! 1_j\ra \Xu_j &= | 0_j \ra.
 \label{ladder-spin-2}
\end{aligned}
\ee
Therefore, $\Xd_j$ and $\Xu_j$ can be deemed as ladder-down and ladder-up pseudo-operators.
Analogous to a spin system, we can also introduce the pseudo-spin-operator $X_j^z$ as
\be \label{define-X-z}
X_{j}^{z} |m_j \rangle = | m_j \ra X_{j}^{z} =  m_j | m_j \ra, \enskip  {\rm for} \; m_j =0, \pm 1.
\ee
The pseudo-operators satisfy the following relations when acted from the left:
\be \label{X-algebra-l}
\{X_{j}^{+}, X_{j'}^{-}\} = \delta_{jj'} X_{j}^{z}, \enskip  [X_{j}^{z}, X_{j'}^{\pm}] = \pm \delta_{jj'} X_{j}^{\pm};
\ee
while for the right action, we have
\be \label{X-algebra-2}
\{X_{j}^{+}, X_{j'}^{-}\} = -\delta_{jj'} X_{j}^{z}, \enskip [X_{j}^{z}, X_{j'}^{\pm}] = \mp \delta_{jj'} X_{j}^{\pm}.
\ee
Here, $[\cdot , \cdot]$ and $\{\cdot , \cdot\}$
represent the commutator and anti-commutator, respectively.

The MAS mapping preserves the exact solution of \Eq{eom-y-1}.
However, such a mapping is intrinsically approximate
because of the finite size of $S_j$. For instance,
some products of AGFs (such as $\eta_{j\tau_1}\eta_{j\tau_2}\eta_{j\tau_3}$)
would become zero during the evolution, which inevitably leads to loss of memory.
Consider another prototypical equation including a convolution integral,
\be
\dot{y} = y \left[D(t)\,\eta_t
+ \int_{0}^t C(t-\tau)\,\baeta_\tau {\rm d}\tau \right], \label{eom-y-2}
\ee
where $C(t)$ and $D(t)$ are c-number functions.
The MAS mapping of \Eq{sub-2} does not recover the exact solution of $\la y \ra$.
Instead, it yields a reasonable approximation; see Ref.~\onlinecite{Han19050601} for details.

\subsection{MAS-SEOM method for open fermionic systems}

With the MAS mapping of \Eq{sub-2}, the stochastic reduced density matrix
of the system $\trhos$ is now defined in the product space $V = V_{_{\rm S}} \otimes S$,
where $V_{_{\rm S}}$ is the subspace of system, and $S = S_{1} \otimes S_{2} \otimes S_{3} \otimes S_{4}$.
In the product space $V$, $\trhos(t)$ can be represented as
\be \label{rho-ex}
\trhos = \sum_{l_1 \in S_1} \sum_{l_2 \in S_2}\sum_{l_3 \in S_3} \sum_{l_4 \in S_4} \trhos^{\,[l_1, l_2, l_3, l_4]},
\ee
with $\trhos^{\,[l_1, l_2, l_3, l_4]}$ being a component
corresponding to the pseudo-Fock-state $|l_1, l_2, l_3, l_4\ra$ of the auxiliary space $S$.
In particular, we choose the initial condition to be
$\trhos(0) = \rho_{_{\rm S}}(0) \otimes |\bm 0 \rangle$,
where $|\bm 0 \rangle = |0_1\ra \otimes |0_2  \ra \otimes |0_3\ra \otimes |0_4\ra$
is the pseudo-vacuum state of $S$.

The SEOM of \Eq{eom-trhos} is cast into the following form:
\begin{align}
\dot{\tilde{\rho}}_{_{\rm S}}
&=  -i [ \Hs, \trhos]
+ e^{-i\pi/4}\hlambda (\hc^\dag\, Y_{1t} + Y_{2t} \,\hc)\trhos \nl
&\quad
+ e^{i\pi/4}\hlambda \trhos(\hc^\dag \,Y_{3t} + Y_{4t}\, \hc),
\label{seom-trhos-2}
\end{align}
where the pseudo-operators $\{Y_{j\tau}\}\,(j=1,\cdots,4)$
are defined by
\be\label{def-y-1}
\begin{split}
	Y_{1t} &\equiv v_{1t} \Xd_1 + \tgm_t, %\Xd_2,
	\quad\   Y_{2t} \equiv v_{2t} \Xu_2 - \tgp_t, %\Xu_1,
	\\
	Y_{3t} &\equiv v_{3t} \Xd_3 - i \tgm_t, %\Xd_2,
	\ \  \,
	Y_{4t} \equiv v_{4t} \Xu_4 + i \tgp_t. %\Xu_1.
\end{split}
\ee
Through the MAS mapping, the memory-convoluted AGFs in \Eq{eom-trhos},
$\{g_t^\pm\}$, are replaced by $\{\tilde{g}^\pm_t\}$ as follows,
\be\label{def-gt-2}
\begin{split}
	\tilde{g}^-_t &=  \lambda^{-1}\big(\,\varphi_{4t} \Xd_4 -i \varphi_{2t} \Xd_2\,\big),\\
	\tilde{g}^+_t &=  \lambda^{-1}\big(\,\varphi_{3t} \Xu_3 -i\varphi_{1t} \Xu_1\,\big),
\end{split}
\ee
where $\{\varphi_j\}\,(j=1,\cdots,4)$ are memory-convoluted c-number noises:
\bea \label{define-varphi-accum}
\varphi_{1, 2}(t) &=& \int_{0}^{t}d\tau \, C^{\pm}(t - \tau)\, v_{1, 2}(\tau), \nonumber \\
\varphi_{3, 4}(t) &=& \int_{0}^{t}d\tau \, [C^{\mp}(t - \tau)]^{*}\, v_{3, 4}(\tau).
\eea
Based on \Eqs{ladder-spin-1} and \eqref{ladder-spin-2},
$\{X^\pm_j\}$ acting on the left or right of $\trhos$ satisfy the following relation:
\be \label{x:rule1}
X^\pm_j \trhos^{\,[l_1, l_2, l_3, l_4]} =
(-1)^{l_1+l_2+l_3+l_4} \trhos^{\,[l_1, l_2, l_3, l_4]}X^\pm_j.
\ee

Equation~\eqref{seom-trhos-2} is termed as the MAS-SEOM,
with which the reduced density matrix of system is obtained by
projecting $\trhos(t)$ to the pseudo-vacuum state of the MAS, and then
taking the stochastic average over all the Gaussian white noises $\{v_{j\tau}\}$ at $0 < \tau < t$:
\be \label{def-avrho}
\brho_{_{\rm S}} =\mathcal{M} \mathcal{P}\big(\trhos\big) = \mathcal{M}\Big(\trhos^{[0,0,0,0]}\Big),
\ee
where $\mathcal{P} \equiv \la \bm 0 |$ denotes the projection to
the pseudo-vacuum state.

Compared with the space of the AGFs $\{\eta_{j\tau},\,\baeta_{j\tau}\}\,(j=1,\cdots,4)$,
the auxiliary space $S$ is spanned by $3^4$ pseudo-Fock-states,
and hence the MAS mapping of \Eq{sub-2} greatly reduces the computational cost.
Consequently, the MAS-SEOM can be employed directly to do stochastic calculations.
For a more general situation in which a multi-level system is coupled to more than
one fermion baths, the MAS mapping is also applicable.
The corresponding details of the MAS-SEOM will be elucidated
in \Sec{sec:sim-HEOM}.

As the result of MAS mapping, the time-dependent AGFs in \Eq{eom-trhos}
are represented by time-dependent c-number fields with
time-independent pseudo-levels.
The stochastic transfers of particles between the system and the baths
is replaced by the exchange of particle or hole from the system to the pseudo-levels.
In the language of quantum chemistry, we can describe the AGFs $\{\eta_{j\tau},\baeta_{j\tau}\}$
by the full configuration-interaction (CI) approach for the
pseudo-levels in the auxiliary space.\cite{Sza96}
Nevertheless, the calculation of full CI is normally unfeasible in practice,
so we truncate the auxiliary space by considering finite excitation configurations.
In principle, the MAS is just an approximation at the single CI level,
and the auxiliary space $S_j$ is much smaller than the full space of AGFs.
For instance, \Eq{seom-trhos-2}
only requires $4$ matrices with the size of $3^4 \times 3^4$ to represent all pseudo-operators, which
can be directly applied to the simulation of fermonic dissipative dynamics.
Although the MAS mapping (single CI) is an approximation for interacting systems,
it leads to highly accurate results; see Ref.~\onlinecite{Han19050601}.
In the next section, we will assess the exactness or non-exactness
of MAS-SEOM by different analytic approaches.

\section{Approximation properties and some important features of MAS-SEOM} \label{sec:limitation}

\subsection{Approximate nature of MAS mapping}

Although the MAS mapping of \Eq{sub-2}
preserves many important properties of the AGFs, such as
$\la \eta_{j\tau} \ra = \la \baeta_{j\tau} \ra = 0$ and
$\la \eta_{j\tau} \baeta_{j'\tau'} \ra = - \la \baeta_{j'\tau'} \eta_{j\tau} \ra = \delta_{jj'}\delta(\tau-\tau')$,
the finite size of MAS inevitably causes the loss of memory.
For instance, some products of AGFs, such as $\eta_{j\tau_1}\eta_{j\tau_2}\eta_{j\tau_3}$,
vanish during the evolution of MAS-SEOM.
In this section, we will scrutinize the difference between
the formally exact SEOM of \Eq{eom-trhos} and
the MAS-SEOM of \Eq{seom-trhos-2}.
In particular, we will focus on how the approximate nature of
the MAS mapping affects the accuracy of the resulting SEOM.
Two analytic approaches will be employed:
the time-dependent perturbation theory approach and the HEOM approach.

\subsection{Time-dependent perturbation theory} \label{sec:perturbation}

In this subsection, we assess the exactness or non-exactness of the MAS-SEOM
by comparing the perturbation expansion of MAS-SEOM with
the counterpart of the original \Sch equation of \Eq{def:total-eom}.
Without loss of generality, we still choose a single-level system
coupled to a fermion bath. The total Hamiltonian is
$H_{_{\rm T}} = H_0 + H_{_{\rm SB}}$ with $H_0 = \Hs + \Hb$.
In the $H_0$-interaction picture, the EOM of total density matrix is
\begin{align} \label{stan-eom}
\dot{\rho}_{_{\rm T}}^{\rm I} &= -i\big[ H_{_{\rm{SB}}}^{\rm I},\,{\rho}_{_{\rm T}}^{\rm I}\big]=-i\mathcal{L}_{_{\rm SB}}^{\rm I}\rho_{_{\rm T}}^{\rm I}.
\end{align}
Here, ${\rho}_{_{\rm T}}^{\rm I} \equiv e^{iH_0 t} {\rho}_{_{\rm T}} e^{-iH_0 t}$
and $\iLsb \equiv e^{iH_0 t} \mathcal{L}_{_{\rm SB}} e^{-iH_0 t}$
are the density matrix and the Liouville operator
in the $H_0$-interaction picture, respectively.
$\Hs = \epsilon \, \hc^\dag \hc$ is the Hamiltonian of the single-level system,
and the system-bath coupling Hamiltonian
$H_{_{\rm SB}}=\kappa^{\frac{1}{2}} (\hc^\dag\hF+\hF^\dag\hc)$ with $\kappa$
being the perturbation strength ($\kappa>0$)
is taken as the perturbation.

We start with \Eq{stan-eom} and construct the deterministic time-dependent
perturbation expansion in ascending order of the parameter $\kappa$, i.e.,
$\bar{\rho}_{_{\rm S}}(t) = \trb ({\rho}_{_{\rm T}}^{\rm I}) = \sum_{n=0} \bar{\rho}_{_{\rm S}}^{(2n)}(t)$.
The $n$th-order response of system reduced density matrix to the perturbation
is expressed as
\begin{align}\label{stan-dyson}
\brhos^{(2n)}
&= {(-1)^{n}}\int_{0}^{t}dt_1\cdots\int_{0}^{t_{2n-1}}dt_{2n}\nl
&\qquad \qquad \times \trb\big[\iLsb(t_1)\cdots \iLsb(t_{2n}) \rho_{_{\rm T}} (0)\big],
\end{align}
where the Liouville operators are arranged in the time-ordered form ($t > t_1 >\cdots>t_{2n-1} > t_{2n}$).
The factorized initial condition $\rho_{_{\rm T}} (0) = \rho_0\,\rhob^{\rm eq}$ is adopted,
with $\rho_0 = \hc \hc^\dagger = 1-\hc^\dagger\hc$.
In the following, we explicitly evaluate the low-order ($n=1,2$) responses
of $\brhos$ based on \Eq{stan-dyson}.

The first-order response of $\brhos$ is
\begin{align}\label{def-rho-1st}
\brhos^{(2)}
&=\kappa\int_{0}^{t}dt_1 \int_{0}^{t_{1}}dt_{2}\, \trb \Big\{
 \big[\hc^\dag(t_2) \rho_0 \hc(t_1)\hF^\dag(t_1) \hF(t_2) \nl
& \quad -\hc(t_1)\hc^\dag(t_2) \rho_0\, \hF^\dag(t_1)\hF(t_2) \nl
& \quad - \rho_0 \hc(t_2) \hc^\dag(t_1)\, \hF^\dag(t_2)\hF(t_1) \nl
& \quad + \hc^\dag(t_1) \rho_0 \hc(t_2)\, \hF^\dag(t_2) \hF(t_1) \big] \rhob^{\rm eq} \Big\}.
\end{align}
By using the definitions of two-time bath correlation functions,
\Eq{def-rho-1st} results in
\begin{align} \label{rho-1st-1}
\brhos^{(2)} &= \kappa \int_{0}^{t}dt_{1} \int_{0}^{t_{1}}dt_{2} \, \Big\{ C^+(t_{1} - t_{2})
\big[ \hc^\dag(t_2) \rho_0 \, \hc(t_1) \nl
&\quad -  \hc(t_1)\hc^\dag(t_2)\,\rho_0 \big] + C^+(t_{2} - t_{1}) \nl
&\quad \times \big[\hc^\dag(t_1) \rho_0\, \hc(t_2) - \rho_0\, \hc(t_2) \hc^\dag(t_1) \big]  \Big\}.
\end{align}
Here, $\hc(\tau) \equiv e^{iH_0 \tau} \hc \, e^{-iH_0 \tau} = e^{i\Hs t} \hc \, e^{-i\Hs t}$,
and $C^\pm(\tau)$ are the bath correlation functions given by \Eq{fdt-1}.

The second-order response of $\brhos$ to perturbation is expressed as
\begin{align}
\brhos^{(4)}
&=\int_{0}^{t}dt_1\cdots\int_{0}^{t_3}dt_4\,\trb\big[\iLsb(t_1)\cdots \iLsb(t_4)\rho_{_{\rm T}} (0)\big] \nl
&= \sum_{\bm \gamma} \brho_{\bm \gamma}^{(4)}, \label{def-rho-2nd}
\end{align}
where ${\bm \gamma}=(\gamma_1,\cdots,\gamma_4)$ with $\gamma_j=\rm{L},\rm{R}$.
$\brhos^{(4)}$ consists of $2^4 = 16$ terms, and $\rm L\,(\rm R)$ means that $H_{_{\rm SB}}^{\rm I}$ acts
on $\rho_{_{\rm T}} (0)$ from left (right).
Each term can be expressed as
\bea \label{def-rho-2nd-1}
\brho_{\bm \gamma}^{(4)}&=& \int_{0}^{t}dt_{1} \ldots \int_{0}^{t_{3}}dt_{4} \, \bar{R}_{\bm{\gamma}}^{(4)}(t_{1}, \ldots, t_{4}),
\eea
where $\{\bar{R}_{\bm{\gamma}}^{(4)}(t_{1}, \ldots, t_{4})\}$ involve
the four-time bath correlation functions. For example,
$\bar{R}_{\rm LLLL}^{(4)}(t_{1}, \ldots, t_{4})$ is expressed as
\begin{align} \label{def-2nd-corr}
\bar{R}_{\rm LLLL}^{(4)}
&=\kappa^2 \, \hc(t_1) \hc^\dag (t_2) \hc(t_3) \hc^\dag(t_4) \rho_0 \nl
&\quad \times
\trb \big[\hF^{\dagger}(t_1) \hF(t_2) \hF^{\dagger}(t_{3})\hF (t_{4})\rhob^{\rm eq}\big] \nl
%
%&=\kappa^2\, \hc(t_1)\hc^\dag (t_2) \hc(t_3) \hc^\dag(t_4) \rhos(0) \nl
%&\quad \times \Big\{\trb [\hF^{\dagger}(t_1)\hF(t_4)\rhob^{\rm eq}] \trb [\hF(t_2)\hF^{\dagger}(t_3)\rhob^{\rm eq}] \nl
%&\quad + \enskip \trb [\hF^{\dagger}(t_1)\hF(t_2)\rhob^{\rm eq}] \trb [\hF^{\dagger}(t_3)\hF (t_4)\rhob^{\rm eq}] \Big\} \nl
&=\kappa^2\, \hc(t_1) \hc^\dag(t_2) \hc(t_3) \hc^\dag(t_4)\,\rho_0
  \big[C^+(t_1-t_2) \nl
&\quad \times C^+(t_3-t_4) + C^+(t_1-t_4) C^-(t_2-t_3)\big].
\end{align}
Here, the second equality makes use of the Gaussian statistical property
that a high-order bath correlation function can be fully expressed
by the two-time correlation functions $C^\pm(\tau)$.
The other 15 members of $\{\bar{R}_{\bm{\gamma}}^{(4)}(t_{1}, \ldots, t_{4})\}$
can be expressed in a similar fashion.

In the following, we start with the MAS-SEOM, and build a perturbation expansion in the
stochastic framework. The results will be compared directly with the above formulas
based on \Eq{stan-eom}.
In the interaction picture of $\Hs$, the MAS-SEOM of \Eq{seom-trhos-2} can be expressed as the following:
\begin{align}
\dot{\tilde{\rho}}_{_{\rm S}}^{\,\rm I}
&=e^{-i\pi/4}\, \hlambda\, \big[\hc^\dag(t)\, Y_{1t} + Y_{2t} \,\hc(t) \big]\, \itrhos \nl
&\quad + e^{i\pi/4}\, \hlambda\, \itrhos\, \big[\hc^\dag(t) \,Y_{3t} + Y_{4t}\, \hc(t)\big] \nl
&=-i\tilde{\mathcal{L}}^{\,\rm I}\, \itrhos,
\label{mas-seom-i}
\end{align}
where $\tLsb\equiv e^{i\Hs t} \tilde{\mathcal{L}}\, e^{-i\Hs t}$ is
the stochastic Liouville operator in the interaction picture,
and the pseudo-operators $\{Y_{j\tau}\}$ are defined by
\be\label{def-iy-1}
\begin{split}
	Y_{1t} &\equiv \kappa^{\frac{1}{2}} \big(v_{1t} \Xd_{1} + \tgm_t\big), %\Xd_2,
	\quad\   Y_{2t} \equiv  \kappa^{\frac{1}{2}}\big(v_{2t} \Xu_{2} - \tgp_t\big), %\Xu_1,
	\\
	Y_{3t} &\equiv  \kappa^{\frac{1}{2}}\big(v_{3t} \Xd_{3} - i \tgm_t\big), %\Xd_2,
	\ \ \
	Y_{4t} \equiv  \kappa^{\frac{1}{2}}\big(v_{4t} \Xu_{4} + i \tgp_t\big).\nonumber %\Xu_1.
\end{split}
\ee

We now take the stochastic Liouvillian as the perturbation, and construct
the perturbation expansion in ascending order of the parameter $\kappa$,
$\itrhos = \sum_{n=0}\trhos^{(2n)}$.
Similar to \Eq{stan-dyson}, the $n$th-order response of
system reduced density matrix is expressed as
\begin{align} \label{def-sDyson}
\trhos^{(2n)}
&={(-1)^n} \int_{0}^{t}dt_1\cdots\int_{0}^{t_{2n-1}}dt_{2n} \, \tLsb(t_1)\,\cdots \nl
&\qquad \qquad\qquad \cdots \tLsb(t_{2n})\, \trhos(0),
\end{align}
where the Liouville operators are also arranged in the time-order form
($t > t_1 >\cdots>t_{2n-1} > t_{2n}$), and the initial condition is
${\trhos}{(0)} = \rho_0 \otimes| \bm 0 \ra$.

The first-order response of $\la \trhos \ra$ is
\begin{align}\label{def-srho-1st}
\la \tilde{\rho}_{_{\rm S}}^{(2)} \ra
&= -\int_{0}^{t}dt_1\int_{0}^{t_1}dt_2 \, \big\la \tLsb(t_1)\,\tLsb(t_2)\,\trhos (0)\big\ra\nl
&= \! \int_{0}^{t} \! dt_1 \! \int_{0}^{t_1} \! dt_2
\Big[\! -i\lambda \hc(t_1)\hc^\dag (t_2) \rho_0 \mathcal{M}\mathcal{P}\big(Y_{2t_1} Y_{1t_2}|\bm 0\ra \big) \nl
&\qquad \qquad +\lambda\,\hc^\dag (t_2)\rho_0\, \hc(t_1) \,
\mathcal{M}\mathcal{P}\big( Y_{1t_2} |\bm 0\ra Y_{4t_1}\big) \nl
&\qquad \qquad +i\lambda\,\rho_0 \, \hc(t_2)\hc^\dag (t_1) \,
\mathcal{M}\mathcal{P}\big(|\bm 0\ra Y_{4t_2}\, Y_{3t_1}\big) \nl
&\qquad \qquad +\lambda\,\hc^\dag (t_1)\rho_0\, \hc(t_2) \,
\mathcal{M}\mathcal{P}\big(Y_{1t_1} |\bm 0\ra Y_{4t_2}\big) \Big] \nl
&= \kappa\,\int_{0}^{t} dt_1 \int_{0}^{t_1} dt_2 \, \Big[ \hc(t_1)\hc^\dag (t_2) \, \rho_0 \nl
&\qquad \times
\mathcal{M}(\varphi_{1t_1}v_{1t_2})\, \mathcal{P}({\Xu_1}\Xd_{1}|\bm 0\ra) \nl
&\qquad + \hc^\dag (t_2) \rho_0\, \hc(t_1) \,
\mathcal{M}(\varphi_{1t_1}v_{1t_2})\, \mathcal{P}(\Xd_1|\bm 0\ra\Xu_1) \nl
&\qquad + \rho_0 \, \hc(t_2)\hc^\dag (t_1) \,
\mathcal{M}(\varphi_{4t_1}v_{4t_2})\, \mathcal{P}(|\bm 0\ra \Xu_4 \Xd_4) \nl
&\qquad +  \hc^\dag (t_1)\rho_0\, \hc(t_2) \,
\mathcal{M}(\varphi_{4t_1}v_{4t_2})\, \mathcal{P}(\Xd_4 |\bm 0\ra \Xu_4) \Big].
%
%&=-\kappa \, \int_{0}^{t}dt_{1} \int_{0}^{t_{1}}dt_{2} %e^{-i\epsilon (t_{1} - t_{2})}
%C^+(t_{1} - t_{2})\hc(t_1)\hc^\dag(t_2) \nl
%&\quad +\kappa \int_{0}^{t}dt_{1} \int_{0}^{t_{1}}dt_{2} %e^{-i\epsilon(t_{1} - t_{2})}
%C^+(t_{1} - t_{2}) \hc^\dag(t_2) \hc(t_1) \nl
%&\quad -\kappa \int_{0}^{t}dt_{1} \int_{0}^{t_{1}}dt_{2} %e^{\, i\epsilon(t_{1} - t_{2})}
%[C^+(t_{1} - t_{2})]^\ast\hc(t_2)\hc^\dag(t_1)  \nl
%&\quad +\kappa \int_{0}^{t}dt_{1} \int_{0}^{t_{1}}dt_{2} %e^{i\epsilon(t_{1} - t_{2})}
%[C^+(t_{1} - t_{2})]^\ast\hc^\dag(t_1) \hc(t_2)
\end{align}
It can be easily verified that the final expression of
$\la \tilde{\rho}_{_{\rm S}}^{(2)} \ra$ in \Eq{def-srho-1st} is identical to
$\brhos^{(2)}$ of \Eq{rho-1st-1}, i.e.,
the MAS-SEOM  exactly recovers the first-order response of $\brhos$.

The second-order response of $\la \trhos \ra$ is
\begin{align}
 \trhos^{(4)}
&=\int_{0}^{t}dt_1\cdots\int_{0}^{t_3}dt_4\, \tLsb(t_1)\cdots\tLsb(t_{4})\,{\trhos}{(0)} \nl
&= \sum_{\bm \gamma} \trho_{\bm \gamma}^{(4)}, \label{def-srho-2nd}
\end{align}
where ${\bm \gamma}=(\gamma_1,\cdots,\gamma_4)$ with $\gamma_j= \rm{L}, \rm{R}$.
Similar to \Eq{def-rho-2nd}, $\trhos^{(4)}$ consists of $16$ terms, and $\rm L\,(\rm R)$
means that the pseudo-operators $\{Y_{j\tau}\}$ acts on
$\trhos$ from left (right).
The statistical average of the second-order response is
$\la \trhos^{(4)} \ra = \sum_{\bm \gamma} \la \trho_{\bm \gamma}^{(4)} \ra$, with
\be \label{def-srho-2nd-1}
\la \trho_{\bm \gamma}^{(4)} \ra = \int_{0}^{t}dt_{1} \cdots \int_{0}^{t_{3}}dt_{4} \,
\la \tilde{R}_{\bm{\gamma}}^{(4)}(t_{1}, \ldots, t_{4}) \ra.
\ee
For example, for $\bm\gamma = ({\rm LLLL})$, we have
\begin{align}\label{def-2nd-scorr}
\la{\tilde{R}}^{(4)}_{{\rm LLLL}}\ra &=
-\lambda^{2}\,\hc(t_1) \, \hc^\dag (t_2) \, \hc(t_3)\, \hc^\dag(t_4)\, \rho_0 \nl
&\quad \times  \mathcal{M} \mathcal{P} \big(\, Y_{2t_1} Y_{1t_2} Y_{2t_3} Y_{1t_4} |\bm 0\ra\big) \nl
&= \kappa^2\, \hc(t_1)\hc^\dag (t_2) \hc(t_3) \hc^\dag(t_4)\, \rho_0
 \Big[\mathcal{M}(\varphi_{1t_1}v_{1t_2}) \nl
&\quad \times
\mathcal{M}(\varphi_{1t_3}v_{1t_4})\, \mathcal{P}(\Xu_1\Xd_1\Xu_1\Xd_1|\bm 0\ra) \nl
&\quad - \mathcal{M}(\varphi_{1t_1}v_{1t_4}) \, \mathcal{M}(\varphi_{2t_2}v_{2t_3}) \nl
&\quad \times \mathcal{P}(\Xu_1\Xd_2\Xu_2\Xd_1|\bm 0\ra) \Big].
%
%&=\kappa^2\hc(t_1)\hc^\dag (t_2) \hc(t_3) \hc^\dag(t_4)\brhos(0)\nl &\quad \times
%\big[ C^+(t_1-t_2) C^+(t_3-t_4)\nl &\quad + C^+(t_1-t_4) C^-(t_2-t_3)\big].
\end{align}
It can be verified that the final expression of $\la{\tilde{R}}^{(4)}_{{\rm LLLL}}\ra$
in \Eq{def-2nd-scorr} is identical to ${\bar{R}}^{(4)}_{{\rm LLLL}}$ of \Eq{def-2nd-corr}.
Likewise, it is found that $\la \tilde{R}_{\bm \gamma}^{(4)}\ra = \bar{R}_{\bm \gamma}^{(4)}$
holds for other $\bm\gamma$, and thus
$\la \trhos^{(4)} \ra = \brhos^{(4)}$.
Therefore, the MAS-SEOM exactly reproduces the first- and second-order
response of $\brhos$.

The discrepancy of the MAS-SEOM from the exact Liouville equation
emerges from the third-order response.
From \Eq{stan-dyson}, the third-order response of $\brhos$ is
a summation of $2^6 = 64$ contributions,
\begin{align} \label{def-rhos-3nd-1}
\brhos^{(6)} &= \sum_{\bm\gamma}  -\int_{0}^{t}dt_{1} \ldots \int_{0}^{t_{5}}dt_{6} \, \bar{R}_{\bm{\gamma}}^{(6)}(t_{1} \ldots, t_{6}) \nl
&= \sum_{\bm\gamma} \brho^{(6)}_{\bm\gamma},
\end{align}
where $\bm\gamma = (\gamma_1, \ldots, \gamma_6)$ with $\gamma_j = \rm{L}, \rm{R}$.
For example, $\bar{R}^{(6)}_{\rm LLLLLL}$ is expressed as
\begin{align} \label{def-3nd-corr}
\bar{R}^{\,(6)}_{\rm LLLLLL} &= \kappa^3\, \hc(t_1)\, \hc^\dag(t_2) \, \hc(t_3)\, \hc^\dag(t_4)\,
 \hc(t_5)\, \hc^\dag(t_6)\, \rho_0 \nl
&\quad \times\trb \big[\hF^{\dagger}(t_1)\hF(t_2)\hF^{\dagger}(t_3)\hF(t_4)
\hF^{\dagger}(t_5)\hF(t_6)\rhob^{\rm eq}\big] \nl
&=  \kappa^3\, \hc(t_1)\, \hc^\dag(t_2) \, \hc(t_3)\, \hc^\dag(t_4)\,
 \hc(t_5)\, \hc^\dag(t_6)\, \rho_0 \nl
&\quad \times \bar{r}_{\rm LLLLLL}^{\,(6)},
\end{align}
where $\bar{r}_{\rm LLLLLL}^{\,(6)}$ is a sixth-order bath correlation function.
Because of the Gaussian statistics, the trace over the bath DOF
for the evaluation of $\bar{r}_{\rm LLLLLL}^{\,(6)}$
can be contracted by the Wick's theorem into $3! = 6$ terms.\cite{Mat92}%%% The Wick theorem
Here, we only select one of them and use it as the reference for the
perturbation analysis based on the MAS-SEOM. The selected term is
\be \label{def-3nd-corr-1}
 \bar{r}_{\rm LLLLLL}^{\,(6)}(\bm t') = C^+(t_1-t_6)C^-(t_2-t_5)C^+(t_3-t_4),
\ee
where the vector ${\bm t'} = \big\{\{t_1,\,t_6\},\,\{t_2,\,t_5\},\,\{t_3,\,t_4\}\big\}$
indicates the pairs of time instants corresponding to the contracted operator pairs.

Based on the MAS-SEOM, the third-order response of $\trhos$ is
\be \label{def-srho-3nd}
\trhos^{(6)}  = %\sum_{\bm \gamma} \trho_{\bm \gamma}^{(6)} =
 - \sum_{\bm \gamma} \int_{0}^{t}dt_{1} \ldots \int_{0}^{t_{3}}dt_{4} \, \tilde{R}_{\bm{\gamma}}^{(6)}(t_{1}, \ldots, t_{6}),
\ee
where $\bm\gamma = (\gamma_1, \ldots, \gamma_6)$ with $\gamma_j = \rm{L}, \rm{R}$.
In particular, the statistical average of $ \tilde{R}_{\rm LLLLLL}^{(6)}(t_{1}, \ldots, t_{6})$ is
\begin{align} \label{def-3nd-scorr}
%\hspace*{-2cm}
\big\la \tilde{R}^{\,(6)}_{\rm{L} \cdots \rm{L}} \big\ra &=
-i \lambda^3 \, \hc(t_1)\, \hc^\dag(t_2)\,
\hc(t_3)\, \hc^\dag(t_4) \, \hc(t_5)\, \hc^\dag(t_6)\, \rho_0\nl
&\quad \times \mathcal{M}\mathcal{P}\big(Y_{2t_1} Y_{1t_2} Y_{2t_3} Y_{1t_4} Y_{2t_5} Y_{1t_6}|\bm 0 \ra\big) \nl
&=  \kappa^3 \, \hc(t_1)\, \hc^\dag(t_2)\, \hc(t_3)\, \hc^\dag(t_4)\,
\hc(t_5)\, \hc^\dag(t_6)\, \rho_0 \nl
&\quad \times \big\la \tilde{r}_{\rm LLLLLL}^{\,(6)} \big\ra.
\end{align}
In relation to \Eq{def-3nd-corr-1}, we explicitly
examine the component
\begin{align}\label{def-3nd-scorr-1}
\big\la \tilde{r}_{\rm LLLLLL}^{\,(6)}(\bm t') \big\ra &= C^+(t_1-t_6)C^-(t_2-t_5)C^+(t_3-t_4)\nl
&\quad \times \mathcal{P}\big( X_{1}^+ X_{2}^- X_{1}^+ X_{1}^- X_{2}^+ X_{1}^-|\bm 0\ra \big) \nl
&= 0.
\end{align}
Here, $X_1^-$ acts on the pseudo-vacuum state $|\bm 0\ra$ twice consecutively,
and thus yields the zero value of $\tilde{r}_{\rm LLLLLL}^{\,(6)}(\bm t')$.
Consequently,
$\la \tilde{r}_{\rm LLLLLL}^{\,(6)}(\bm t') \ra \neq \bar{r}_{\rm LLLLLL}^{\,(6)}(\bm t')$.
Similar discrepancies are also found for other components of $\{\tilde{r}_{\bm \gamma}^{(6)}\}$ --
altogether 36 out of 384 components are different.
Therefore, the third-order response of system reduced density matrix
obtained from the MAS-SEOM, $\la \trhos^{(6)}\ra$, does not fully recover $\brhos^{(6)}$
of the original Liouville equation. It is thus clear that the finite size of
auxiliary space $S_j$ indeed leads to loss of memory for the reduced system dynamics.

In this subsection, based on the time-dependent perturbation theory,
we have demonstrated that the MAS-SEOM reproduces the exact
reduced system dynamics up to the second-order response.
The discrepancy of higher-order response is clearly
ascribed to the finite size of the MAS.
However, from the perturbation theory it is hard to tell
how significantly the discrepancy will affect the accuracy of
the numerical results of MAS-SEOM.
In the following subsection,
we will address this issue by making connection to the HEOM formalism.

\subsection{Equivalence between the MAS-SEOM and a simplified HEOM method}
\label{sec:sim-HEOM}
%

%In this subsection, we further assess how the reduced size of MAS
%affects the exactness of MAS-SEOM by making conection
%to the fermionic HEOM formalism.\cite{Jin08234703,Han18234108}
%

For the convenience of analysis, we consider again a single-level
system coupled to a fermion bath.
Moreover, the number of pseudo-operators $\{X^\pm_j\}$ involved
in the MAS-SEOM of \Eq{seom-trhos-2} are halved by assuming
$X_{3}^\pm = X_{1}^\pm$ and $X_{4}^\pm = X_{2}^\pm$.\cite{Han19050601}
Accordingly, the memory-convoluted fields $\tilde{g}^\pm_t$ are simplified as
\begin{align} \label{def-gt-3}
	\tilde{g}^-_\tau &=  \lambda^{-1}\big(\,\varphi_{4\tau}-i \varphi_{2\tau}\,\big)\Xd_2, \nl
	\tilde{g}^+_\tau &=  \lambda^{-1}\big(\,\varphi_{3\tau}-i\varphi_{1\tau}\,\big)\Xu_1,
\end{align}
where $\{\varphi_{j\tau}\}$ are given in \Eq{define-varphi-accum},
and the resulting MAS-SEOM still has the form of \Eq{seom-trhos-2}.

We now examine the HEOM associated with the MAS-SEOM.
By unraveling the bath correlation functions via \Eq{cexp}, we have
$\tilde{g}^\pm_t = \sum_{m=1}^M \tilde{g}^\pm_m(t)$, with
\begin{align} \label{def-mas-g}
\tgm_{m} &= \lambda^{-1} \! \int_{0}^t d\tau
\big[\! -i A^{-}_{m} v_{2\tau} + \big(A^{+}_{m} \big)^\ast v_{4\tau} \big] e^{\gamma_{m}^{-}(t-\tau)}  X_{2}^-,  \nl
\tgp_{m} &= \lambda^{-1} \! \int_{0}^t d\tau
\big[\! -i A^{+}_{m} v_{1\tau} + \big(A^{-}_{m} \big)^\ast v_{3\tau} \big] e^{\gamma_{m}^{+}(t-\tau)}  X_{1}^+, \nl
\end{align}
An $(I+J)$th-tier ADO of the HEOM is defined by
\be \label{def-mas-ado-3}
\trho_{{m_1} \cdots m_I n_1 \cdots n_J}^{(-\cdots-+\cdots+)} \equiv
(e^{i\pi/4} \hlambda)^{I+J}\, \big\la \tgm_{m_1} \cdots \tgm_{m_I} \trhos\,\tgp_{n_1} \cdots \tgp_{n_J} \big\ra,
\ee
where $\la \cdots \ra \equiv \mathcal{M}\mathcal{P}(\cdots)$,
and the subscript of the ADO is determined by the
sequence of $\{\tilde{g}_{m_i}^-\}$ and $\{\tilde{g}_{n_j}^+\}$
at the left and right of $\trhos$.

It is immediately recognized that the ADO is zero by definition
if the right-hand side of \Eq{def-mas-ado-3} involves two or more
identical pseudo-operators $X_j^\sigma$.
This is because of the finite size of the MAS $S_j$,
so that a pseudo-level can accommodate at most one particle or hole,
and thus $\mathcal{P}[(X^\sigma_{j})^p\, \trhos] =
\mathcal{P}[\,\trhos(X^\sigma_{j})^p] = 0$ for $p > 1$.
For instance,
$\trho_{m_1 m_2}^{(--)}=\la\tgm_{m_1}\tgm_{m_2}\trhos\ra = 0$,
because $\mathcal{P}(X_2^- X_2^- \trhos)=0$.
Likewise, the ADOs $\{\trho_{n_1n_2}^{(++)},\, \trho_{mrn}^{(---)}, \,\trho_{mrn}^{(--+)}, \ldots\}$
are all zero.
By referring to the full HEOM of \Eq{def:ado-1},
these ADOs have a common feature: they involve two or more
generating functionals ($\mathcal{B}_m^\sigma$) that have the same $\sigma$
(differ only in the index $m$).
Such ADOs are termed as the interference ADOs.\cite{Han18234108}

Evidently, the HEOM corresponding to the MAS-SEOM of
\Eq{seom-trhos-2} is a finite hierarchy which terminates automatically at the second tier.
The only nonzero ADOs in the hierarchy are $\{\trho^{(0)},\trho_m^{(+)},\trho_m^{(-)},\trho_{mn}^{(-+)}\}$.
Presuming that $X_{j}^\sigma$ commutes with $\hc$ and $\hc^\dag$
and using the equality $ \la X_{j}^\sigma \trhos \ra = -\la\,\trhos X_{j}^\sigma \ra$,
it can be proved that the EOM of any nonzero ADO has the same form of \Eq{heom-1}.
Such a finite hierarchy with all the interference ADOs excluded is termed
as the simplified HEOM (sim-HEOM).
Its detailed derivation as well as the main features are provided in Ref.~\onlinecite{Han18234108}.

To better understand the correspondence between
the MAS mapping and the zero interference ADOs,
we further consider a more general case in which
a system of $N_\nu$ levels is coupled to $N_\alpha$ fermion baths.
The corresponding MAS-SEOM is
\begin{align} \label{seom-trhos-gen-1}
\dot{\tilde{\rho}}_{_{\rm S}}
&=  -i [ \Hs, \trhos]
+ \hlambda\,\sum_{\nu=1}^{N_\nu} \sum_{\alpha=1}^{N_\alpha} \Big[
e^{\frac{-i\pi}{4}}
\big(\hc^\dag_\nu\, Y_{1\nu\alpha t} + Y_{2\nu\alpha t}\, \hc_\nu\big)\trhos \nl
&\quad
+ e^{\frac{i\pi}{4}}\,
 \trhos \big( \hc^\dag_\alpha \,Y_{3\nu\alpha t} + Y_{4\nu\alpha t}\, \hc_\nu \big)
 \Big].
\end{align}
Here, $\nu$ and $\alpha$ label the system levels and the fermion baths, respectively.
The pseudo-operators $\{Y_{j\nu\alpha\tau}\}\,(j=1,\cdots,4)$
are defined by
\be\label{def-y-gen-1}
\begin{split}
	Y_{1\nu\alpha t} &\equiv v_{1\nu\alpha t} \Xd_{1\nu} + \tgm_{\nu\alpha t}, %\Xd_2,
	\quad  Y_{2\nu\alpha t} \equiv v_{2\nu\alpha t} \Xu_{2\nu} - \tgp_{\nu\alpha t}, %\Xu_1,
	\\
	Y_{3\nu\alpha t} &\equiv v_{3\nu\alpha t} \Xd_{1\nu} - i \tgm_{\nu\alpha t}, %\Xd_2,
	\  \
	Y_{4\nu\alpha t} \equiv v_{4\nu\alpha t} \Xu_{2\nu} + i \tgp_{\nu\alpha t}. %\Xu_1.
\end{split}
\ee
If there is no cross-correlation between any two system levels via the bath,
we have $\tilde{g}^\pm_{\nu\alpha t} = \sum_{m=1}^M \tilde{g}^\pm_{\nu\alpha m}(t)$
by unraveling the bath memory, with
\begin{align} \label{def-mas-cgf}
\tgm_{\nu\alpha m} &= {\lambda}^{-1} \int_{0}^t d\tau \,
\big[ -i A^{-}_{\nu\alpha m}\, v_{2\nu\alpha\tau}
 + (A^{+}_{\nu\alpha m})^\ast\, v_{4\nu\alpha\tau} \big] \nl
&\qquad \qquad \qquad \times e^{\gamma_{\nu\alpha m}^{-}(t-\tau)} X_{2\nu}^-, \nl
\tgp_{\nu\alpha m} &= {\lambda}^{-1} \int_{0}^t d\tau\,
\big[ -i A^{+}_{\nu\alpha m} v_{1\nu\alpha\tau} + (A^{-}_{\nu\alpha m})^\ast\, v_{3\nu\alpha\tau} \big] \nl
&\qquad \qquad \qquad \times e^{\gamma_{\nu\alpha m}^{+}(t-\tau)} X_{1\nu}^+.
\end{align}
The Gaussian white noises satisfy
$\mathcal{M}(v_{j\nu\alpha\tau} v_{j'\nu' \alpha'\tau'}) =
\delta_{jj'}\delta_{\nu\nu'}\delta_{\alpha\alpha'}\delta(\tau-\tau')$.

In the following, we use $p$ and $q$ to denote the multi-component
index $(\nu\alpha m)$. Based on the MAS-SEOM of \Eq{seom-trhos-gen-1},
an $(I+J)$th-tier ADO is constructed by
\be \label{def-mas-ado-4}
\trho_{{p_1} \cdots p_I q_1 \cdots q_J}^{(-\cdots-+\cdots+)} \equiv
(e^{i\pi/4} \hlambda)^{I+J}\, \big\la \tgm_{p_1} \cdots \tgm_{p_I} \trhos\,\tgp_{q_1} \cdots \tgp_{q_J} \big\ra.
\ee
In the framework of HEOM the same ADO is defined as
\begin{align}
\rho_{p_1 \ldots p_I q_1 \cdots q_J}^{(-\cdots-+\cdots+)} &\equiv
\int \mD \bm\bpsi \, \mD\bm\psi\, \mD \bm\bpsi' \, \mD\bm\psi'\,
e^{iS_{\!f}} \mathcal{F}_{_{\rm FV}} e^{-iS_{\rm b}} \nl
&\qquad \times \mathcal{B}^{-}_{p_I} \cdots \mathcal{B}^{-}_{p_1}
\mathcal{B}^{+}_{q_J} \cdots \mathcal{B}^{+}_{q_1}\, \rhos(0),
\label{def:ado-2}
\end{align}
with the generating functionals given by
\begin{align} \label{def-gen-b}
{{\mathcal{B}}}^{-}_{\nu\alpha m} &=
-i \int_{0}^{t} d\tau \,\big[ A^{-}_{\nu\alpha m } \psi_{\nu \tau}
- \big(A^{+}_{\nu\alpha m}\big)^\ast \,\psi'_{\nu \tau} \big]e^{\gamma_{\nu\alpha m}^{-}(t-\tau)}, \nl
{{\mathcal{B}}}^{+}_{\nu\alpha m} &=
-i\int_{0}^{t} d\tau \,\big[ A^{+}_{\nu\alpha m} \bpsi_{\nu \tau}
- \big(A^{-}_{\nu\alpha m}\big)^\ast \,\bpsi'_{\nu \tau} \big]e^{\gamma_{\nu\alpha m}^{+}(t-\tau)}. \nl
\end{align}
If the ADO defined by \Eq{def:ado-2} includes two or more generating
functionals (${\mathcal{B}}^{\sigma}_{\nu\alpha m}$) that have the
same $\sigma$ and $\nu$ (differ only in $\alpha$ or $m$),
it belongs to the interference ADOs.
Based on the MAS-SEOM of \Eq{seom-trhos-gen-1},
such an interference ADO must have a zero value because
$\mathcal{P}[(X^\sigma_{j\nu})^p \trhos] = \mathcal{P}[\trhos(X^\sigma_{j\nu})^p] = 0$
for $p > 1$.
Therefore, the MAS-SEOM of \Eq{seom-trhos-gen-1} is formally
equivalent to the sim-HEOM in which all the interference ADOs are excluded.\cite{Han18234108}

Alternatively, if $X^\sigma_{j\nu}$ is replaced by $X^\sigma_{j\nu\alpha}$
in \Eqs{def-y-gen-1} and \eqref{def-mas-cgf}, the resulting MAS-SEOM
is formally equivalent to the so-called sim-HEOM-$\alpha$ formalism,\cite{Han18234108}
in which an interference ADO involves two or more generating
functionals (${\mathcal{B}}^{\sigma}_{\nu\alpha m}$) that have the
same $\sigma, \alpha$ and $\nu$ (differ only in the index $m$).
Apparently, the sim-HEOM-$\alpha$ formalism is less approximate than the sim-HEOM,
because in the former a smaller number of interference ADOs are
excluded from the hierarchy.

From the above analysis, it is clear that the MAS mapping itself
is intrinsically approximate, as it may cause loss of
memory (or interference information)
if the particle transfer event occurs consecutively
through a same dissipation mode.
Because the MAS-SEOM and the sim-HEOM (or the sim-HEOM-$\alpha$)
are formally equivalent, they share the following
common features:\cite{Han18234108,Han19050601}

(i) They yield the exact reduced dynamics,
if the bath correlation functions $C^\pm(t)$
have the form of a single exponential function.
This is obvious because the index $m$ would become redundant,
and hence there is no interference ADO
in the original full HEOM.

(ii) They yield the exact single-particle properties
for any non-interacting system.
This is because for non-interacting systems
the HEOM truncated at the second tier
already yield the exact reduced single-particle density matrix,\cite{Jin08234703,Han18234108}
and the interference ADOs have no influence on the latter, i.e.,
$\trs[\hc^\dag \hc\, \trho_{mn}^{(++)}]=\trs[\hc^\dag \hc\,\trho_{mn}^{(--)}]=0$.
Therefore, the interference ADOs can be safely omitted
from the full hierarchy without affecting the exactness
of the resulting single-particle properties.

(iii) For general interacting systems, they are in principle approximate,
and the interference ADOs are supposedly important for the
quantitative description of strong correlation effects
such as the Kondo phenomena.\cite{Hew93}
Indeed, the discrepancies between the results of sim-HEOM and those of
the full HEOM have been demonstrated in Ref.~\onlinecite{Han18234108}.
Nevertheless, as shown in Ref.~\onlinecite{Han19050601} and in our paper~II,\cite{Ari19}
in many cases the MAS-SEOM can still provide reasonably
or even remarkably accurate predictions
for the dissipative dynamics of an interacting system.

\section{Concluding remarks} \label{sec:summary}

To summarize, in this paper we first derive a rigorous SEOM
for describing the dissipative dynamics of fermionic open systems.
The SEOM of \Eq{eom-trhos} is formally exact and is equivalent
to the rigorous fermionic HEOM formalism.
However, such an SEOM is numerically unfeasible because of
the difficulty in modeling g-numbers.

We then propose a MAS mapping scheme with which the time-dependent
g-numbers are mapped to time-dependent c-numbers with a set of pseudo-levels.
This leads to the establishment of a numerically feasible SEOM, termed as the MAS-SEOM.
The MAS-SEOM is found to be equivalent to a simplified version of HEOM
in which the interference ADOs are absent.
The practical implementation and numerical results of the MAS-SEOM
are to be presented in our succeeding paper.\cite{Ari19}

To have an overview of the MAS-SEOM formalism
as well as its potential applicability,
we summarize the advantages and limitations of the present MAS-SEOM in Table~\ref{HEOM-SEOM},
with the present HEOM method listed as a reference.

\begin{table}[t]
	%\caption{Fermionic HEOM vs SEOM}
%	\vspace{10pt}
	\centering
	\begin{tabular}{lcc}
		\hline % \hline
		\thead[l]{}                     & {\small HEOM}   & {\small MAS-SEOM} \\
		\hline
		exactness (non-interacting system)     & Y       & Y         \\
		exactness (interacting system)         & Y       & N         \\
        short-time dynamics                    & Y       & Y          \\
		long-time dynamics          &Y       &N                \\
		stationary state            &Y       &N       \\
		correlated initial state    &Y       &N       \\
		ultra-low temperature       &N       &Y       \\
		massive parallel computation     &Y       &Y       \\
		\hline %\hline
	\end{tabular}
	\caption{An overview of various aspects of the present fermionic HEOM method and the MAS-SEOM method.
        ``Y'' means the method possesses the specific feature or is feasible for the specific situation,
        while ``N'' means the method does not possess the feature or is unfeasible for the situation.
		See the main text in \Sec{sec:summary} for details. \label{HEOM-SEOM}}
\end{table}

The MAS-SEOM shares the common features of the sim-HEOM methods.
For non-interacting systems, the MAS-SEOM yields the
exact dissipative dynamics because all the single-particle properties
of the system are preserved.
In contrast, for interacting systems, the MAS-SEOM method is
intrinsically approximate, because the size of MAS is too small to
record the time ordering of all the environmental fluctuations.
Nevertheless, as shown in Refs.~\onlinecite{Han19050601,Ari19},
remarkably accurate results have been obtained by the MAS-SEOM method.

In the MAS-SEOM formalism, the number of stochastic variables --
the Gaussian white noises keeps increasing as the time evolution proceeds.
Consequently, as will be demonstrated in our paper II,\cite{Ari19}
usually a large number of trajectories are required to achieve
an ample sampling of these stochastic variables, so as to
obtain numerically converged reduced dynamics.
Therefore, the present MAS-SEOM method is very efficient in
the transient dynamics regime, but is much less efficient
for the study of long-time dissipative dynamics.

With the HEOM method, the stationary state (thermal equilibrium state
or non-equilibrium steady state) can be obtained either by propagating
the HEOM to the asymptotic long-time limit,
or by solving the hierarchically coupled linear equations
resulted from the stationary condition.\cite{Li12266403}
However, neither of these two approaches is currently available
for the MAS-SEOM method.
A related issue is that so far only the decoupled initial state
has been considered for the MAS-SEOM. It remains unclear how to formulate
the MAS-SEOM with a correlated initial state,
which is important for many practical purposes.
The imaginary-time SEOM is a potentially promising approach,\cite{Moi12115412,Moi15094108,Ma15094107}
but related works for fermionic environments have not been reported.
Much effort is needed in this direction.

While the numerical aspects of MAS-SEOM are to be
discussed in Ref.~\onlinecite{Ari19},
we would like to point out here that, unlike the HEOM method,
the MAS-SEOM does not require explicit unraveling
of the bath correlation functions, and hence
the cost of computer memory is trivial compared to the HEOM method.
Therefore, the MAS-SEOM is particularly favorable
for exploring the dissipative dynamics at ultra-low temperatures,
which is still challenging for the present HEOM method.
Furthermore, parallel computing techniques have been
applied to the HEOM method.\cite{Kre112166,Str122808,Kre144045,Tsu153859,Kra181779}
Massive parallelization is expected to be very easy for the MAS-SEOM,
since the quantum trajectories are mutually independent and equally weighted.

Finally, because the MAS mapping is intrinsically approximate,
for some interacting open systems,
such as the strongly correlated quantum impurity systems,
the numerical accuracy of the present MAS-SEOM might not be satisfactory.
Even for these systems, the MAS-SEOM provides a valuable foundation for
the future development of more sophisticated stochastic QDTs.

%\section{ACKNOWLEDGMENTS}
\acknowledgments
Support from the Ministry of Science and Technology of China
(Grants No.\ 2016YFA0400900 and No.\ 2016YFA0200600),
the National Natural Science Foundation of China
(Grants No.\ 21973086, No.\ 21573202, No.\ 21633006, No.\ 21973036 and
No.\ 21903078), and the Ministry of Education of China (111 Project Grant No.\ B18051)
is gratefully acknowledged.
V.Y.C. was supported by the U.S. Department of Energy, Office of Science, Basic Energy Sciences,
Materials Sciences and Engineering Division, Condensed Matter Theory Program.

%%%%%%%%%%%%%%%%%%%%%%

%\bibliographystyle{aip}
%\bibliographystyle{aiptit}
%\bibliography{bibrefs}

\end{document}